\documentclass[fleqn,usenatbib,useAMS]{mnras}

\usepackage{graphicx}	
\usepackage{amsmath}	
\usepackage{multicol}   
\usepackage{pdflscape}	
\usepackage{multirow}
\usepackage{booktabs}
\usepackage{siunitx}
\usepackage{float}

\usepackage[T1]{fontenc}
\usepackage{ae,aecompl}

\usepackage{newtxtext,newtxmath}

\title[Kinematic analysis of the TNG50 galaxies]
{Kinematic analysis of the TNG50 Milky Way and Andromeda analogues}

\author[S. I. Denyshchenko et al.]{
S. I. Denyshchenko$^{1}$\thanks{Contact e-mail: sofiia.denyshchenko@gmail.com (SID)},
D. V. Odynets$^{1}$\thanks{Contact e-mail: odynets.d.v@gmail.com (DVO)},
V. S. Akhmetov$^{1,2,3}$\thanks{Contact e-mail: akhmetovs@gmail.com (VSA)},
A. M. Dmytrenko$^{1,3}$,
V. P. Khramtsov$^{1,3}$,
\newauthor
I. I. Kyrylenko$^{1}$,
I. B. Vavilova$^{3}$
and K. O. Miroshnychenko$^{1}$
\\
$^{1}$Institute of Astronomy of V. N. Karazin Kharkiv National University, Svobody sq. 4, 61022, Kharkiv, Ukraine \\
$^{2}$INAF--Osservatorio Astrofisico di Torino, Via Osservatorio 20, Pino Torinese, 10025, Italy \\
$^{3}$Main Astronomical Observatory of the NAS of Ukraine, 27 Akademika Zabolotnoho St., Kyiv 03143, Ukraine
}

\date{Accepted XXX. Received YYY; in original form ZZZ}
\pubyear{\the\year{}}

\begin{document}
\label{firstpage}
\pagerange{\pageref{firstpage}--\pageref{lastpage}}

\maketitle
\begin{abstract}
We investigate whether Ogorodnikov--Milne (O--M) parameters can serve as kinematic tracers of spiral structure in four Milky Way- and Andromeda-like disc galaxies from the TNG50-1 simulation at $z=0$. Stellar particles are analysed in disc-aligned reference frames, with velocities referenced to the mass-weighted mean stellar velocity within one stellar half-mass radius. Local O--M fits on a regular grid yield centroid velocities and the components of the velocity-gradient tensor, characterising local deformation and rotation. We compare their spatial distributions with stellar surface-density and relative overdensity maps and examine the influence of parameter uncertainties. The galaxies exhibit diverse coherent kinematic patterns. Features selected using the generalised Oort constants $A$, $B$, and $C+K$ display partial correspondence with stellar overdensities, depending on the galaxy and the adopted centroid-selection criteria. For TNG50-456326, a galaxy-specific selection enables us to fit four logarithmic spiral branches and determine their geometrical parameters. Their morphology suggests two principal arms and two secondary branches. We compare the fitted spirals with stellar- and gas-mass overdensity and radial velocity maps and examine the spatial relationship between the identified stellar structures and the gas distribution. Qualitative similarities to the morphology and kinematic patterns of the Milky Way make TNG50-456326 a promising candidate for further investigation as a kinematic analogue. Our results support the use of local O--M parameters as complementary diagnostics of non-axisymmetric disc structure, while highlighting the need for galaxy-specific selection criteria.
\end{abstract}



\section{Introduction}
\label{sec:introduction}

The stellar velocity field of a disc galaxy is shaped by both its axisymmetric gravitational potential and its non-axisymmetric structures. Spiral arms, bars, rings, warps, and external perturbations can produce coherent variations in the radial, azimuthal, and vertical motions of stars \citep[e.g.][]{LinShu1964, SellwoodBinney2002, DobbsBaba2014}. Consequently, spatially resolved stellar velocity fields and their gradients provide information that is complementary to the stellar-density distribution and can be used to characterise large-scale kinematic structures and investigate their association with non-axisymmetric features in galactic discs \citep{GaiaCollaboration2018}.

The Milky Way is the only large disc galaxy for which detailed stellar phase-space information is available over an extended volume. The astrometric and spectroscopic data provided by the \textit{Gaia} mission make it possible to study the three-dimensional stellar velocity field over a substantial part of the Galactic disc \citep{GaiaCollaboration2016, GaiaCollaboration2023}. However, because the Sun is located within the Galactic disc, reconstructing its global structure is challenging. Interstellar extinction, spatially varying observational completeness, and uncertainties in stellar distances further complicate the mapping of its morphology and kinematics.

The Ogorodnikov--Milne (O--M) model describes the local stellar velocity field through a first-order Taylor expansion \citep{Ogorodnikov1932, Milne1935, Chandrasekhar1942, Ogorodnikov1965}. The corresponding velocity-gradient tensor can be decomposed into a symmetric deformation tensor and an antisymmetric local-rotation tensor. The diagonal components of the symmetric tensor characterise local expansion or contraction, while its off-diagonal components describe shear deformation. This decomposition provides a framework for examining local variations in the velocity field and identifying spatially coherent kinematic patterns.

The O--M model has previously been applied to observational data to investigate the kinematics of the Milky Way. Using red giants and subgiants from \textit{Gaia} EDR3, \citet{Fedorov2021} determined 12 kinematic parameters for stellar systems distributed along the Galactic centre--Sun--Galactic anticentre direction. Subsequently, \citet{Fedorov2023} constructed two-dimensional maps of the centroid velocities, the O--M parameters, and the spatial derivatives of the stellar velocity field across a substantial part of the Galactic plane. These maps revealed extended and spatially coherent kinematic structures that cannot be fully described by an axisymmetric velocity field.

The deformation tensor derived from \textit{Gaia} DR3 data was also analysed by \citet{Dmytrenko2023}, who investigated the vertex coordinates of stellar systems whose centroids are distributed across the Galactic plane. This study showed that the deformation tensor can provide information about the orientation and non-axisymmetric properties of local stellar velocity distributions. Building on this approach, \citet{Denyshchenko2024} investigated the possible connection between kinematic maps and the spiral structure of the Milky Way. In that work, the centroids of spherical stellar regions with radial deformation velocities consistent with zero within their uncertainties were used as kinematic tracers of spiral arms. The resulting distribution of tracers was used to identify the Scutum--Centaurus, Sagittarius--Carina, Perseus, and Norma--Outer arms, as well as the Local arm, and to determine their pitch angles and Galactocentric positions.

Using a large sample of high-luminosity stars from \textit{Gaia} DR3, \citet{Akhmetov2024} applied a Taylor expansion of the stellar velocity field up to second order and constructed spatial maps of the velocity components and their derivatives, revealing complex kinematic substructures and pronounced departures from axisymmetry across the Galactic disc. \citet{Fedorov2026Rotation} constructed the Milky Way circular-velocity curve out to $R\simeq20$~kpc using \textit{Gaia} DR3 RGB stars and examined its azimuthal dependence, while \citet{Griv2026} estimated the angular rotation speed of the Galactic spiral pattern from the kinematics of main-sequence OB, A, and F stars. These studies show that spatially resolved stellar kinematics can constrain both the axisymmetric rotation of the Milky Way and the dynamical properties of its non-axisymmetric structure.

Although these results suggest that local deformation parameters can help identify kinematic patterns associated with large-scale structure in the Milky Way, their interpretation is limited by the fact that the Galaxy is observed from within the disc, which complicates the independent reconstruction of its global morphology. Cosmological simulations offer a complementary way to assess this approach: the three-dimensional positions and velocities of stellar particles are available directly, allowing the kinematic patterns identified by the O--M model to be compared with the intrinsic stellar surface-density distribution and its overdense features.

The TNG50 cosmological magnetohydrodynamical simulation \citep{Pillepich2019, Nelson2019, SotilloRamos2022, Pillepich2024} is the highest-resolution run of the IllustrisTNG project\footnote{\url{https://www.tng-project.org/}} and provides a detailed view of the structural, morphological, and kinematic properties of galaxies within a cosmological volume of approximately $50$ Mpc on a side. \citet{Pillepich2019} investigated the structural and kinematic evolution of star-forming galaxies in TNG50. \citet{Zana2022} developed a kinematic decomposition of TNG50 galaxies into cold and warm discs, bulges, and stellar haloes. \citet{Emami2022} studied the radial dependence of stellar velocity anisotropy in Milky Way+Andromeda (MW/M31) TNG50 galaxies, while \citet{SotilloRamos2023} investigated disc flaring and its relation to vertical stellar kinematics in MW/M31-like systems. Finally, \citet{Pillepich2024} presented a sample of 198 MW/M31 analogues, characterised their structural and environmental properties, and analysed the stellar and gaseous velocity dispersions of selected galaxies. More broadly, \citet{Vavilova2026a} proposed a hierarchical framework for identifying Milky Way near twins based on structural, dynamical, environmental, and evolutionary diagnostics. This framework was subsequently applied to four nearby galaxies by \citet{Vavilova2026b}, highlighting the importance of combining complementary diagnostics when assessing their similarity to the Milky Way.

In this work, we apply the Ogorodnikov--Milne model to four MW/M31-like galaxies selected from the TNG50 simulation \citep{Pillepich2024} to investigate the relationship between local stellar kinematics and disc structure. We construct two-dimensional maps of the local velocity-gradient components and compare the structures identified using the generalised Oort constants $A$, $B$, and $C+K$ with the stellar surface-density and relative overdensity distributions. Our aim is to assess the potential of these kinematic diagnostics to trace spiral and other non-axisymmetric structures and to examine how their spatial correspondence depends on the galaxy and the adopted selection criteria. We also seek to identify candidate kinematic analogues of the Milky Way by comparing the simulated velocity fields and their local gradients with those observed in our Galaxy.

The simulated galaxies provide an opportunity to investigate the relation between such kinematic signatures and the underlying stellar morphology under controlled conditions. Unlike the Milky Way, their stellar morphology and density distribution can be measured directly. It should be emphasised that spiral structure in both observed and simulated galaxies is generally more complex than an ideal logarithmic spiral. Galactic discs may contain arm bifurcations, branches, spurs, feathers, and shorter inter-arm features, while the main arms themselves may be patchy or interrupted \citep{Hou2021, SellwoodMasters2022}. Additional substructures, such as spurs and feathers, may extend from the main arms into the inter-arm regions \citep{DobbsBonnell2006, Mandowara2022}. Such features can connect different arm segments or appear as secondary structures with geometries different from those of the primary spiral pattern.

This paper is organised as follows. Section~\ref{sec:data} describes the TNG50 simulation data, the selection of the four analysed galaxies and their stellar samples, the construction of the disc-aligned reference frames, and the stellar-mass overdensity maps. Section~\ref{sec:om} presents the O--M model and its application to the simulated stellar velocity fields. Section~\ref{sec:kinematic-results} presents the resulting centroid-velocity fields and the symmetric deformation and local-rotation components, and examines their relation to stellar overdensities through the selected kinematic tracers. It also presents the logarithmic spiral parameters derived for TNG50-456326 using galaxy-specific selection criteria and compares the fitted spirals with gas and stellar-mass overdensity maps. The spatial distributions of the numbers of stellar particles contributing to the local kinematic solutions are presented in Appendix~\ref{app:starsAmount}, while the corresponding uncertainty maps are shown in Appendix~\ref{app:parameter_errors}. The sensitivity of the kinematic selection in TNG50-456326 to the adopted local fitting radius is examined in Appendix~\ref{app:rloc_sensitivity}. Finally, the main conclusions are summarised in Section~\ref{sec:conclusions}.


\section{Simulation data and sample selection}
\label{sec:data}

\subsection{The TNG50 simulation}
\label{sec:tng50}

The IllustrisTNG project is a suite of cosmological magnetohydrodynamical simulations developed to study galaxy formation and evolution in a $\Lambda$CDM Universe \citep{Nelson2018, Pillepich2018, Springel2018}. The simulations were performed with the moving-mesh code \textsc{Arepo} \citep{Springel2010} and include gravity, magnetohydrodynamics, chemical enrichment, and feedback from stars and active galactic nuclei \citep{Weinberger2017, Pillepich2018}. Among the three main IllustrisTNG volumes, TNG50 provides the highest spatial and mass resolution and is therefore particularly suitable for studies of the internal structure and kinematics of individual galaxies \citep{Pillepich2019, Nelson2019}.

In this study, we use the highest-resolution realisation of TNG50, referred to as TNG50-1. TNG50-1 follows the evolution of dark matter, gas, stars, supermassive black holes, and magnetic fields in a periodic cosmological volume with a comoving side length of 51.7~Mpc. The initial conditions contain $2160^{3}$ dark-matter particles and $2160^{3}$ gas cells. The mass of an individual dark-matter particle is approximately $4.5\times10^{5}~\mathrm{M}_{\odot}$, while the target baryonic mass resolution is $8.5\times10^{4}~\mathrm{M}_{\odot}$ \citep{Pillepich2019, Nelson2019}.

The collisionless gravitational softening length for stellar and dark-matter particles is approximately 288~pc at $z=0$, while the adaptive spatial resolution of gas cells in star-forming regions is typically 70--140~pc \citep{Pillepich2019}. This combination of mass and spatial resolution allows Milky Way-mass galaxies to be resolved with millions of stellar particles while preserving their cosmological environments.

Our analysis is based on snapshot 99 of TNG50-1, corresponding to the present epoch ($z=0$). Dark-matter haloes are identified using the Friends-of-Friends (FoF) algorithm \citep{Davis1985} with a linking length of 0.2 times the mean separation between particles. Gravitationally bound substructures are subsequently identified using the \textsc{Subfind} algorithm \citep{Springel2001, Dolag2009}. Each bound substructure is assigned a unique Subhalo ID within the snapshot. The most massive subhalo of a FoF halo is classified as its central galaxy, while the remaining subhaloes are classified as satellites. The Subhalo IDs used throughout this work refer to the TNG50-1 group catalogue at snapshot 99.

\subsection{Data selection and disc-frame construction}
\label{sec:selected_galaxies}


\begin{table*}
\centering
\caption{Properties of the four selected TNG50-1 galaxies
at snapshot 99 ($z=0$). The stellar sample satisfies
$0<R\leq30\,\mathrm{kpc}$ and $|Z|\leq1\,\mathrm{kpc}$
in the disc frame.}
\label{tab:galaxies_r30}

\setlength{\tabcolsep}{3pt}
\renewcommand{\arraystretch}{1.15}

\resizebox{\textwidth}{!}{%
\begin{tabular}{lrrrrrrrrrrr}
\hline
\hline
Subhalo ID
& $\log_{10}(M_{\rm sub}/M_\odot)$
& $M_{\star,30}$
& $R_{1/2,\star}$
& SFR
& $N_\star$
& $N_{\rm disc}$
& $f_{\rm disc}$
& $\mathrm{med}(|Z|)$
& $\langle v_R\rangle$
& $\mathrm{med}(|v_\vartheta|)$
& $\langle v_Z\rangle$ \\
&
&
$(10^{11}M_\odot)$
& (kpc)
& $(M_\odot\,\mathrm{yr}^{-1})$
&
&
&
& (kpc)
& $(\mathrm{km\,s^{-1}})$
& $(\mathrm{km\,s^{-1}})$
& $(\mathrm{km\,s^{-1}})$ \\
\hline

388544
& 12.673 & 1.3630 & 11.104 & 8.10
& 3136208 & 969904 & 0.309
& 0.334 & -0.12 & 167.27 & -0.20 \\

432106
& 12.440 & 1.3929 & 10.939 & 9.67
& 2550617 & 1129514 & 0.443
& 0.217 & 0.02 & 197.20 & 0.26 \\

456326
& 12.322 & 1.0578 & 12.477 & 8.85
& 1932429 & 946672 & 0.490
& 0.326 & 2.04 & 177.38 & -0.03 \\

467415
& 12.334 & 1.0964 & 5.813 & 12.60
& 1927474 & 1175309 & 0.610
& 0.293 & 0.09 & 214.11 & -0.48 \\

\hline
\end{tabular}%
}

\vspace{2mm}

\begin{minipage}{\textwidth}
\footnotesize
$M_{\rm sub}$ is the total gravitationally bound subhalo mass. $M_{\star,30}$ is the stellar mass within 30 kpc, and $R_{1/2,\star}$ is the stellar half-mass radius. SFR is the instantaneous subhalo star formation rate. $N_\star$ is the total number of genuine stellar particles, $N_{\rm disc}$ is the number in the selected disc sample, and $f_{\rm disc}=N_{\rm disc}/N_\star$. The height and velocity statistics refer to the selected sample in the disc frame oriented using \texttt{temet.meanAngMomVector}, with star-forming gas and stellar particles within one stellar half-mass radius. Velocities are measured relative to the mass-weighted mean velocity of genuine stellar particles within a sphere of one stellar half-mass radius, without a vertical cut. The quantities $\langle v_R\rangle$ and $\langle v_Z\rangle$ are arithmetic means over the selected disc particles, without mass weighting.
\end{minipage}

\end{table*}

\begin{figure*}
\centering
\includegraphics[width=0.9\textwidth]{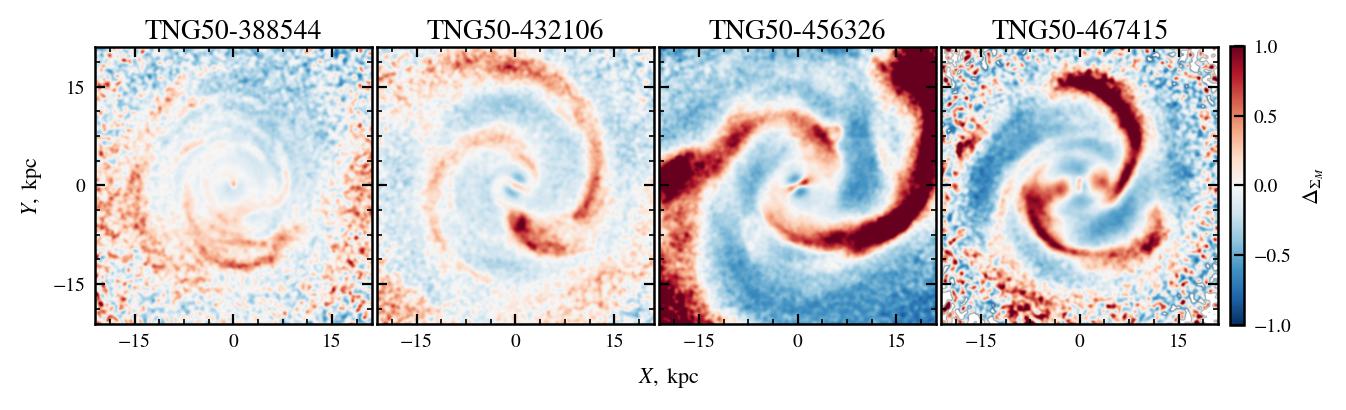}
\caption{
Relative stellar-mass surface-density maps in the disc plane for the four selected TNG50-1 MW/M31-like galaxies at snapshot 99 ($z=0$). The local stellar-mass surface density, $\Sigma_{M,\mathrm{loc}}(X,Y)$, was obtained from the mass-weighted stellar-particle distribution smoothed with a Gaussian kernel of width $\sigma=0.25$~kpc. The axisymmetric background, $\Sigma_{M,0}(R)$, was derived from the azimuthally averaged radial profile in annuli of width $\Delta R=0.4$~kpc.
}
\label{fig:density_maps}
\end{figure*}

The galaxies analysed in this work were selected from the public TNG50 Milky Way and Andromeda analogue catalogue. The catalogue contains 198 MW/M31-like galaxies at $z=0$ and provides their Subfind IDs, stellar and halo masses, star formation rates, structural parameters, morphological classifications, environmental properties, and satellite populations \citep{SotilloRamos2022, Pillepich2024}.

The baseline MW/M31 analogue sample is defined using criteria based on stellar mass, morphology, and environment. Galaxies are required to have stellar masses in the range $10^{10.5}<M(<30\,\mathrm{kpc})/\mathrm{M}_{\odot}<10^{11.2}$, disc-dominated morphologies identified through the axis-ratio criterion $c/a < 0.45$, and relatively isolated environments with no neighbouring galaxy of comparable stellar mass within 500~kpc and host halo masses below $10^{13}\,\mathrm{M}_{\odot}$ \citep{SotilloRamos2022}. Particular attention was given to galaxies containing a sufficiently large number of stellar particles to ensure reliable spatial sampling and robust determination of local velocity gradients using the O--M model. We also prioritised systems with well-developed stellar discs and identifiable spiral structures.

We analysed four galaxies from the TNG50-1 simulation identified by their Subfind IDs 388544, 432106, 456326, and 467415. The stellar and gas cutouts, together with the required subhalo properties, were obtained through the IllustrisTNG public data API \citep{Nelson2019Data}. For each galaxy, we obtained stellar particles (\texttt{PartType4}) and gas cells (\texttt{PartType0}), together with the subhalo position, bulk velocity, and stellar half-mass radius. In the IllustrisTNG data structure, \texttt{PartType4} contains both genuine stellar particles and wind-phase particles; we therefore retained only particles with $\mathrm{GFM\_StellarFormationTime}>0$ as genuine stars. The main properties of the four selected galaxies and their stellar samples are summarised in Table~\ref{tab:galaxies_r30}.

The selected galaxies span a broad range of disc properties. The TNG50 \texttt{SubhaloSFR} values, which represent instantaneous star formation rates summed over the gas cells of each subhalo, range from 8.1 to 12.6~$M_\odot\,\mathrm{yr^{-1}}$ for the selected galaxies. For comparison, observational estimates place the present-day star formation rate of the Milky Way at approximately $1.6$--$1.9~M_\odot\,\mathrm{yr^{-1}}$ \citep{Chomiuk2011,Licquia2015}. This comparison is affected by differences between observational SFR indicators and the TNG50 \texttt{SubhaloSFR} definition. Although the selected galaxies can serve as Milky Way analogues in terms of their discs properties, they may not be close analogues in terms of their current star formation activity. The absolute median azimuthal velocities of the selected stellar discs
span approximately $167$--$214~\mathrm{km\,s^{-1}}$.

To emphasise non-axisymmetric stellar structures, we constructed relative stellar-mass overdensity maps following approaches commonly used to trace spiral structure in both numerical simulations and observational studies of the Galactic disc \citep{Grand2016,Poggio2021}. The stellar-mass surface-density maps were constructed from the same genuine stellar-particle sample satisfying $0<R\leq30~\mathrm{kpc}$ and $|Z|\leq1~\mathrm{kpc}$.

The local stellar-mass surface density, $\Sigma_{M,\mathrm{loc}}(X,Y)$, is defined on a regular $(X,Y)$ grid with a pixel size of $0.1$~kpc. Stellar particles contribute according to their masses, and the resulting two-dimensional distribution is smoothed with a Gaussian kernel of width $\sigma_{\mathrm{loc}}=0.25$~kpc. The reference surface density is taken to be axisymmetric and depends only on Galactocentric radius. It is defined from the azimuthally averaged local surface density in concentric annuli of width $\Delta R=0.4$~kpc. The resulting radial profile is smoothed with a one-dimensional Gaussian kernel of width $0.25$~kpc and interpolated over the two-dimensional grid, defining the axisymmetric background surface density $\Sigma_{M,0}(R)$. The relative stellar-mass overdensity is then defined as

\begin{equation}
\Delta_{\Sigma_M}(X,Y)
= \frac{\Sigma_{M,\mathrm{loc}}(X,Y)} {\Sigma_{M,0}(R)} -1. \notag
\end{equation}

Positive values of $\Delta_{\Sigma_M}$ correspond to an excess of stellar mass relative to the azimuthally averaged density at the same radius, while negative values indicate a deficit. This representation suppresses the dominant radial density gradient and emphasises non-axisymmetric structures in the stellar disc.

The stellar-mass overdensity maps in Fig.~\ref{fig:density_maps} reveal substantial morphological diversity among the selected galaxies. Despite belonging to the same broad MW/M31 analogue class, the systems differ in the strength, radial extent, and coherence of their stellar structures. Some galaxies exhibit relatively well-defined large-scale spiral features, whereas others show more fragmented or flocculent patterns. These morphological differences provide a useful reference for interpreting the corresponding stellar velocity fields (Fig.~\ref{fig:VR_Vtheta_VZ}) and the spatial distributions of the local deformation and rotation parameters (Figs.~\ref{fig:M11_M22_M33}, \ref{fig:M12_M13_M23}, and \ref{fig:omega1_omega2_omega3}).

The orientation of each disc was determined using \texttt{meanAngMomVector} from the \texttt{temet} analysis package\footnote{\url{https://github.com/dnelson/temet}}. This function selects star-forming gas cells with $\mathrm{StarFormationRate}>0$ and stellar particles with $\mathrm{GFM\_StellarFormationTime}\geq0$ within a sphere of one catalogue stellar half-mass radius centred on the Subfind potential minimum. The selected elements contribute their mass-weighted angular momenta, and the mean angular-momentum vector is normalised to obtain the disc axis, $\hat{\boldsymbol{L}}$.

For the stellar kinematic analysis, the velocity reference was defined as the mass-weighted mean peculiar velocity of genuine stellar particles with $\mathrm{GFM\_StellarFormationTime}>0$ within a sphere of one catalogue stellar half-mass radius centred on the Subfind potential minimum:

\begin{equation}
\boldsymbol{v}_{\mathrm{ref}}
=
\frac{\sum_{i\in\mathcal{S}}m_i\boldsymbol{v}_i}
     {\sum_{i\in\mathcal{S}}m_i}, \notag
\end{equation}
where $\mathcal{S}$ denotes this inner stellar sample. No vertical cut was applied when calculating $\boldsymbol{v}_{\mathrm{ref}}$. This reference replaces the catalogue subhalo bulk velocity for the exported stellar kinematics.

Positions were centred on the Subfind potential minimum, corrected for periodic boundaries, and converted to physical units. The reference velocity $\boldsymbol{v}_{\mathrm{ref}}$ was subtracted from the stellar peculiar velocities. The rotation matrix obtained using \texttt{temet.rotationMatrixFromVec} was then applied to the centred positions and relative velocities, initially placing $\hat{\boldsymbol{L}}$ along the positive $Z$-axis. We verified that the matrix was orthonormal and right-handed. To adopt the same coordinate convention as in our previous Milky Way kinematic studies \citep{Fedorov2023,Akhmetov2024}, the resulting frame was subsequently rotated by $180^\circ$ about the $X$-axis. Thus, in the final reference frame the disc angular-momentum vector is directed along the negative $Z$-axis. This choice ensures that the signs of the azimuthal velocity and the corresponding O--M parameters are directly comparable with those used in our previous Galactic analyses. In this frame, cylindrical coordinates and velocities were calculated as

\begin{equation}
R=\sqrt{X^2+Y^2},
\qquad
\theta=\operatorname{atan2}(Y,X), \notag
\end{equation}
\begin{equation}
v_R=\frac{XV_X+YV_Y}{R},
\qquad
v_\theta=\frac{XV_Y-YV_X}{R}. \notag
\end{equation}

The Hubble-flow term included in the \texttt{temet} angular-momentum routine was not added to the exported stellar velocities.

For the subsequent O--M analysis, we selected genuine stellar particles that satisfy $0<R\leq30~\mathrm{kpc}$ and $|Z|\leq1~\mathrm{kpc}$. The same radial and vertical limits were applied to all four galaxies. The resulting catalogue for each galaxy contains particle IDs, stellar masses, disc-frame Cartesian positions and velocities, and the derived cylindrical coordinates and velocities. Positions are expressed in physical kiloparsecs, velocities in $\mathrm{km\,s^{-1}}$, and masses in $M_\odot$.


\section{Ogorodnikov--Milne model}
\label{sec:om}

After transforming all stellar particles into the individual disc-aligned reference frame of each galaxy, the local stellar velocity field is analysed using the Ogorodnikov--Milne (O--M) kinematic model. Although originally developed for stellar motions in the Milky Way \citep{Ogorodnikov1932, Ogorodnikov1965}, the mathematical framework of the model is independent of the underlying stellar system. Here, we apply the same formalism to stellar particles in the simulated TNG50 galaxies.

The disc-aligned reference frame is centred on the Subfind potential minimum, and stellar velocities are measured relative to the reference velocity $\boldsymbol{v}_{\mathrm{ref}}$ defined in Sect.~\ref{sec:selected_galaxies}. Local stellar systems are centred at different positions within this galaxy-centred frame and therefore generally have non-zero coordinates and centroid velocities.

Local stellar systems are defined at the nodes of a regular Cartesian grid in the plane of each galaxy, with a separation of $\Delta_{\mathrm{grid}}=0.1~\mathrm{kpc}$ between adjacent nodes. For each node in the grid, stellar particles within a spherical region of radius $r_{\mathrm{loc}}=1.0~\mathrm{kpc}$ centred on the node are used to determine local kinematic parameters. Since the radius of the local region is larger than the grid spacing, neighbouring regions overlap; the corresponding local solutions are therefore not statistically independent, but the overlap provides a densely sampled spatial representation of the velocity field and the O--M parameters. A local solution is accepted only when the region contains at least $N_{\min}=5$ stellar particles. This criterion primarily affects the low-density outer regions of the stellar discs, where fewer particles are available within a local region. The corresponding spatial distributions of the number of stellar particles contributing to the local kinematic solutions are shown in Fig.~\ref{fig:starsAmount}.

For every accepted local region, the stellar velocity field is represented by a first-order Taylor expansion around the position $\boldsymbol{r}_0=(X_0,Y_0,Z_0)$ of the corresponding local stellar system, following the approach described in \citet{Fedorov2023}:

\begin{equation}
V_i(\boldsymbol{r})
=
V_i(\boldsymbol{r}_0)
+
\sum_{j=1}^{3}
\left.
\frac{\partial V_i}{\partial x_j}
\right|_{\boldsymbol{r}_0}
(x_j-x_{j,0}). \notag
\label{eq:local-taylor-results}
\end{equation}
where $x_1=X$, $x_2=Y$, and $x_3=Z$, and $V_{i,0}\equiv V_i(\boldsymbol{r}0)$. Here, $V{i,0}$ denotes the $i$-th component of the centroid velocity of the local stellar region relative to the reference velocity $\boldsymbol{v}_{\rm ref}$, while $\partial V_i/\partial x_j$ are the components of the local velocity-gradient tensor evaluated at $\boldsymbol{r}_0$.

The 12 unknown parameters of the local velocity field are determined simultaneously using the least-squares method. The uncertainties of the fitted parameters are derived from the covariance matrix of the least-squares solution. These comprise the three centroid velocities, $V_{X,0}$, $V_{Y,0}$, and $V_{Z,0}$, and the nine spatial derivatives$\partial V_i/\partial x_j$, with $i,j=1,2,3$. The resulting velocity-gradient tensor is decomposed into antisymmetric and symmetric parts,

\begin{equation}
\frac{\partial V_i}{\partial x_j}
=
M^{-}_{ij}
+
M^{+}_{ij},
\end{equation}
where
\begin{equation}
M^{-}_{ij}
=
\frac12
\left(
\frac{\partial V_i}{\partial x_j}
-
\frac{\partial V_j}{\partial x_i}
\right)
\label{eq:antisymmetric-tensor-results}
\end{equation}
describes the local rotational component of the velocity field, while
\begin{equation}
M^{+}_{ij}
=
\frac12
\left(
\frac{\partial V_i}{\partial x_j}
+
\frac{\partial V_j}{\partial x_i}
\right)
\label{eq:symmetric-tensor-results}
\end{equation}
is the symmetric deformation component of the velocity field.

The diagonal components of $M^{+}$ characterise local expansion or contraction along the coordinate axes, whereas the non-diagonal components describe shear deformation in the corresponding coordinate planes. Since $M^{+}$ is symmetric, it contains six independent components, while the antisymmetric tensor $M^{-}$ contains three independent components. The deformation and local-rotation components are expressed in units of $\mathrm{km\,s^{-1}\,kpc^{-1}}$.


\section{Local kinematic fields and spiral structure}
\label{sec:kinematic-results}

\begin{figure*}
\centering
\includegraphics[width=0.81\textwidth]{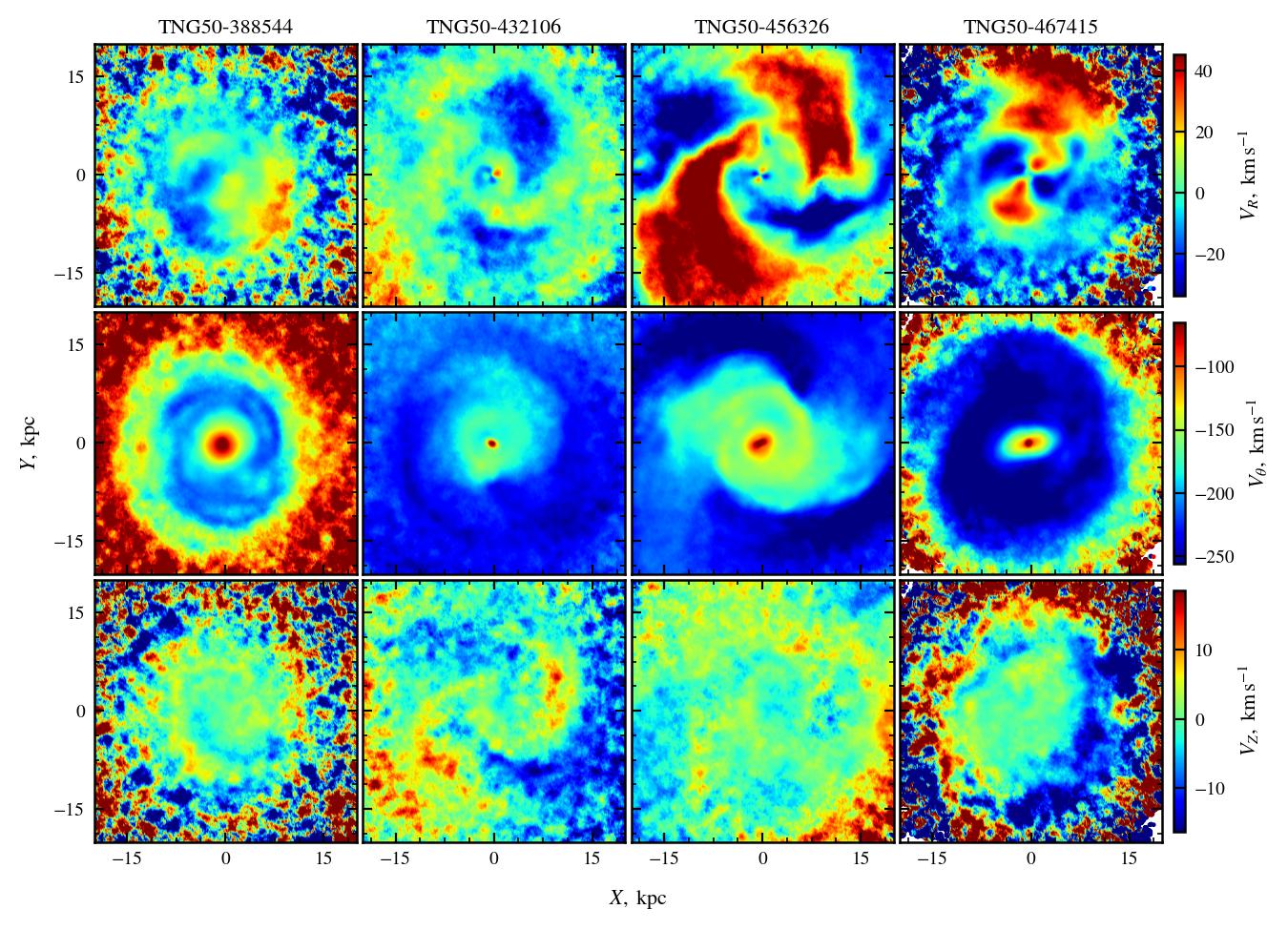}
\caption{Centroid-velocity fields of the stellar component for the selected TNG50-1 galaxies. From top to bottom, the rows show the radial velocity $V_R$, azimuthal velocity $V_\theta$, and vertical velocity $V_Z$, respectively. Each column corresponds to one galaxy listed in Table~\ref{tab:galaxies_r30}.}
\label{fig:VR_Vtheta_VZ}
\end{figure*}

\begin{figure*}
\centering
\includegraphics[width=0.91\textwidth]{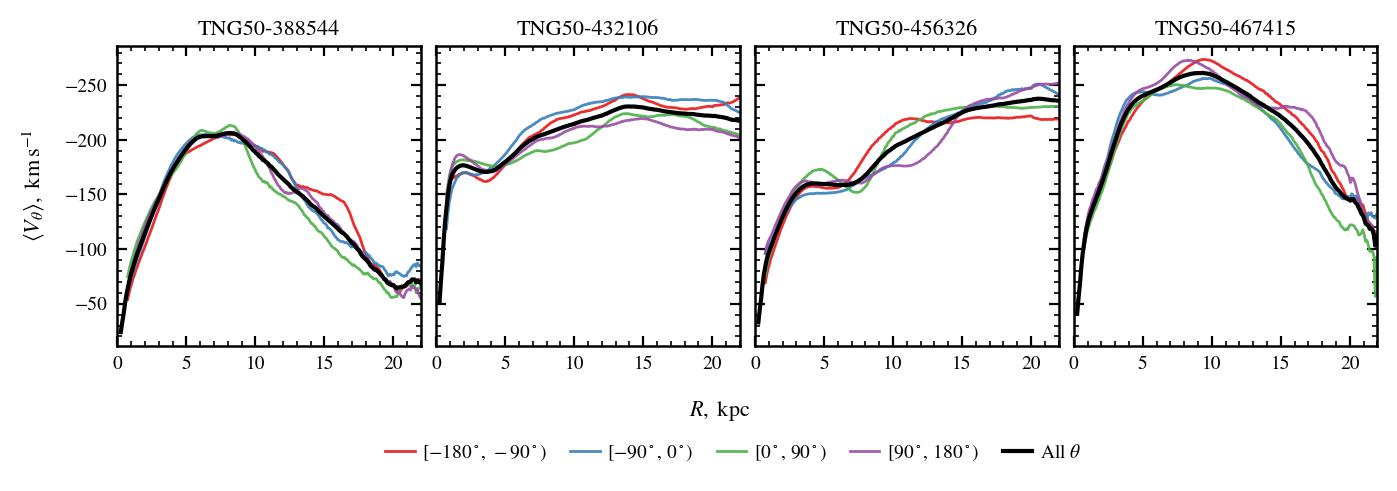}
\caption{
Radial profiles of the mean azimuthal velocity for the selected TNG50-1 galaxies. The black solid curves show $\langle V_\theta\rangle$ calculated over the full azimuthal range, while the coloured curves represent the corresponding profiles in four $90^{\circ}$ azimuthal sectors. The differences between the sector profiles illustrate the azimuthal variations of the stellar rotation field and departures from a purely axisymmetric velocity distribution. All profiles were calculated in radial bins of width $\Delta R=0.1$~kpc.
}
\label{fig:Vth}
\end{figure*}

\begin{figure*}
\centering
\includegraphics[width=0.81\textwidth]{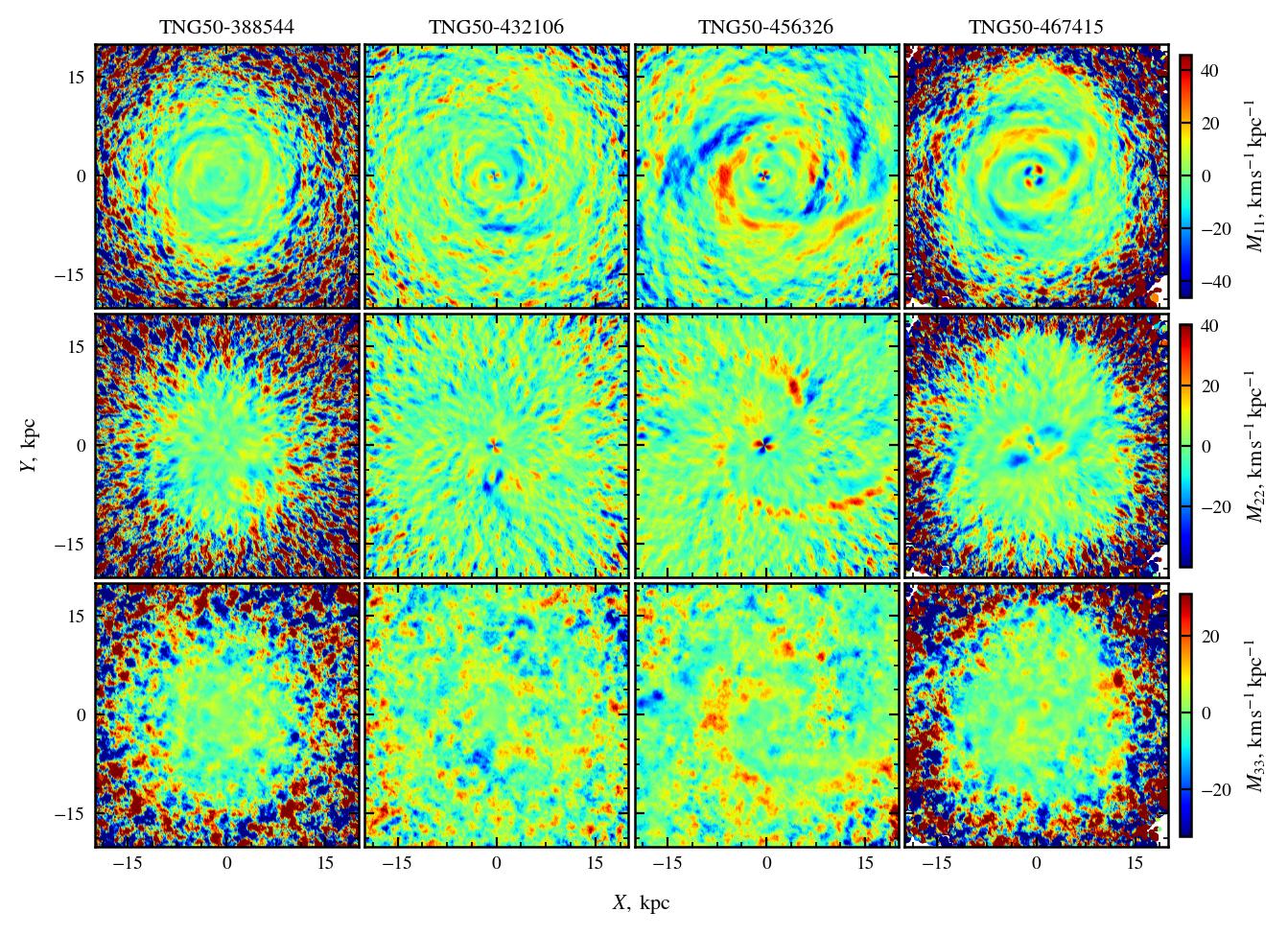}
\caption{
Centroid maps of the stellar velocity-gradient tensor diagonal components $M_{11}^{+}$, $M_{22}^{+}$, and $M_{33}^{+}$ for the selected TNG50-1 galaxies. From top to bottom, the rows show $M_{11}^{+}$, $M_{22}^{+}$, and $M_{33}^{+}$, respectively. Each column corresponds to one galaxy listed in Table~\ref{tab:galaxies_r30}.
}
\label{fig:M11_M22_M33}
\end{figure*}

\begin{figure*}
\centering
\includegraphics[width=0.81\textwidth]{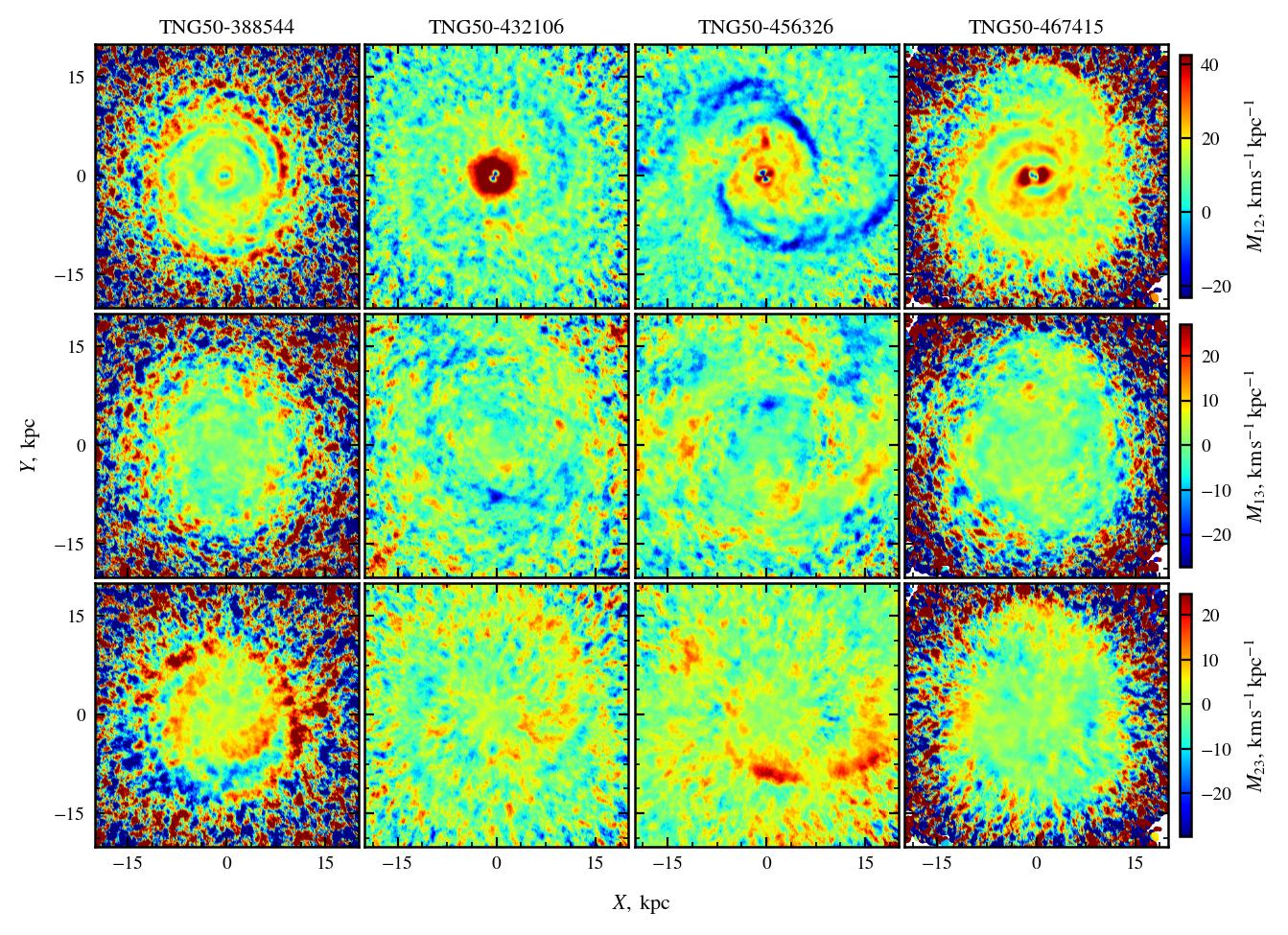}
\caption{
Centroid maps of the stellar velocity-gradient tensor off-diagonal components $M_{12}^{+}$, $M_{13}^{+}$, and $M_{23}^{+}$ for the selected TNG50-1 galaxies. From top to bottom, the rows show $M_{12}^{+}$, $M_{13}^{+}$, and $M_{23}^{+}$, respectively. Each column corresponds to one galaxy listed in Table~\ref{tab:galaxies_r30}.
}
\label{fig:M12_M13_M23}
\end{figure*}

\begin{figure*}
\centering
\includegraphics[width=0.81\textwidth]{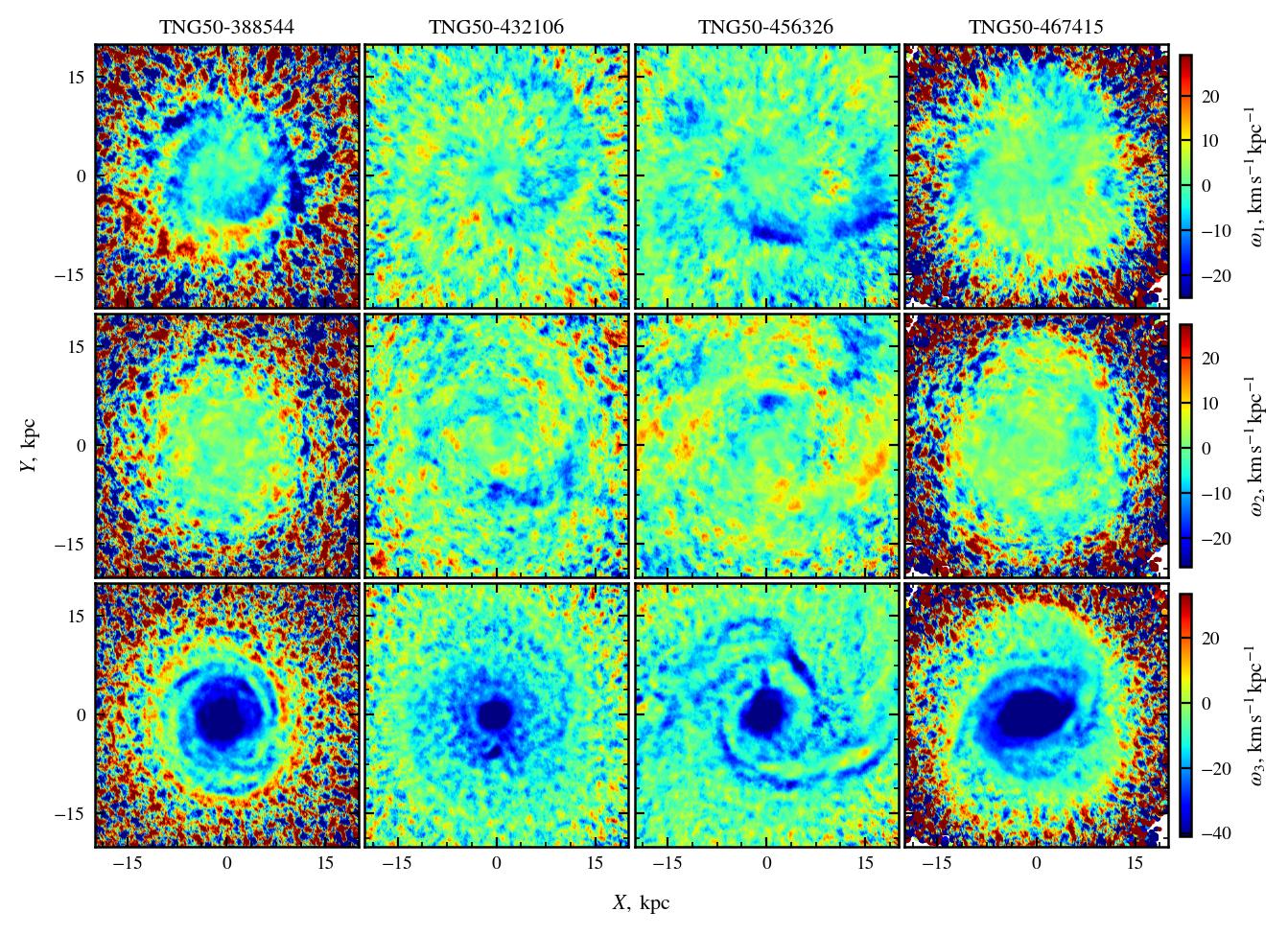}
\caption{
Centroid maps of the local-rotation components $\omega_1$, $\omega_2$, and $\omega_3$ for the selected TNG50-1 galaxies. From top to bottom, the rows show $\omega_1$, $\omega_2$, and $\omega_3$, respectively. Each column corresponds to one galaxy listed in Table~\ref{tab:galaxies_r30}.
}
\label{fig:omega1_omega2_omega3}
\end{figure*}

The O--M analysis provides local centroid velocities and velocity-gradient tensors at each accepted grid position. The centroid velocities describe the local mean stellar motion, while the velocity-gradient tensor characterises its spatial variation.

We examine the spatial distributions of the O--M model parameters and their uncertainties to assess their potential as kinematic tracers of spiral structure. To this end, we compare the spatial patterns in each parameter map with the stellar surface-density distribution, considering their correspondence with spiral features, spatial coherence, and statistical significance. The near-zero $M_{11}^{+}$ criterion proposed by \citet{Denyshchenko2024} is examined as one candidate diagnostic alongside features in the other model parameters.

\subsection{Centroid velocity fields}
\label{sec:centroid-velocities}

The centroid-velocity fields of the selected TNG50-1 galaxies are shown in Fig.~\ref{fig:VR_Vtheta_VZ}. The azimuthal velocity $V_\theta$ shows the dominant rotational motion of the stellar discs, whereas the radial and vertical components exhibit substantially more complex spatial structures.

Radial velocity maps show coherent regions of positive and negative $V_R$ on scales of several kiloparsecs. In several galaxies, these regions form extended or curved patterns rather than a simple axisymmetric distribution, indicating coherent radial streaming motions. The strongest and most extended radial structures are visible in TNG50-456326 and TNG50-467415, while the other systems show weaker or more fragmented radial patterns.

The azimuthal-velocity fields are comparatively smooth and are dominated by the global rotation of the stellar discs. The radial behaviour of $V_\theta$ differs among the galaxies, both in amplitude and in the characteristic scale over which the rotation curve changes. TNG50-388544, and TNG50-467415 show pronounced spatial gradients in $V_\theta$, whereas TNG50-432106, and TNG50-456326 display a comparatively more uniform azimuthal-velocity distribution over substantial parts of their disc.

To examine the radial and azimuthal variations of the stellar rotation in more detail, Fig.~\ref{fig:Vth} shows the mean azimuthal-velocity profiles calculated over the full azimuthal range and separately in four $90^{\circ}$ sectors. The azimuthally averaged profiles trace the overall radial behaviour of the stellar rotation, whereas the sector profiles reveal local departures from axisymmetry. In several galaxies, the sector curves remain close to the mean profile over a substantial radial range, while in others they show systematic offsets and local deviations. These differences indicate that a single azimuthally averaged rotation curve does not always provide a complete description of the stellar velocity field and that the rotational kinematics can vary significantly with azimuth as shown by \citet{Fedorov2026Rotation}. Both the overall radial behaviour and the spatial variations of $V_\theta$ are qualitatively consistent with the large-scale azimuthal-velocity field derived for the Milky Way from Gaia DR3 by \citet{Fedorov2023,Akhmetov2024}.

As summarised in Table~\ref{tab:galaxies_r30}, the mean vertical velocities of the selected stellar discs remain close to zero, although their spatial distributions are considerably more complex. The vertical-velocity maps show regions of both positive and negative $V_Z$ with characteristic spatial scales of several kiloparsecs. The most inhomogeneous distributions are observed in TNG50-432106 and TNG50-467415.  TNG50-388544 and TNG50-456326 show predominantly low vertical velocities, with values remaining close to zero over substantial parts of their stellar discs, although weak local variations are still present. These differences may be related to the individual morphology and dynamical state of each galaxy. Interactions with dwarf satellites, disc warps, recent accretion events, or minor mergers may also contribute to the observed vertical-velocity asymmetries \citep[e.g.][]{Gomez2013, Poggio2021, Semczuk2025}. However, $V_Z$ maps alone cannot uniquely determine their origin; this requires comparison with the galaxies' merger histories, satellite orbits, stellar-density distributions, and evolution across successive simulation snapshots.

The amplitudes, spatial scales, and morphologies of their radial, azimuthal, and vertical velocity structures differ substantially, reflecting differences in their morphological and dynamical states. The spatial behaviour of all three centroid-velocity components, $V_R$, $V_\theta$, and $V_Z$, is qualitatively consistent with the kinematic maps of the Milky Way presented by \citet{Fedorov2023,Akhmetov2024}. In both the simulated galaxies and the \textit{Gaia} DR3-based maps, the azimuthal component is dominated by global disc rotation, while the radial and vertical components exhibit lower-amplitude spatial variations associated with departures from purely axisymmetric motion. This similarity supports the use of the selected TNG50 galaxies as numerical laboratories for testing kinematic diagnostics developed from observations of the Milky Way.

\subsection{Symmetric deformation tensor}
\label{sec:deformation-results}

\begin{figure*}
\centering
\includegraphics[width=0.85\textwidth]{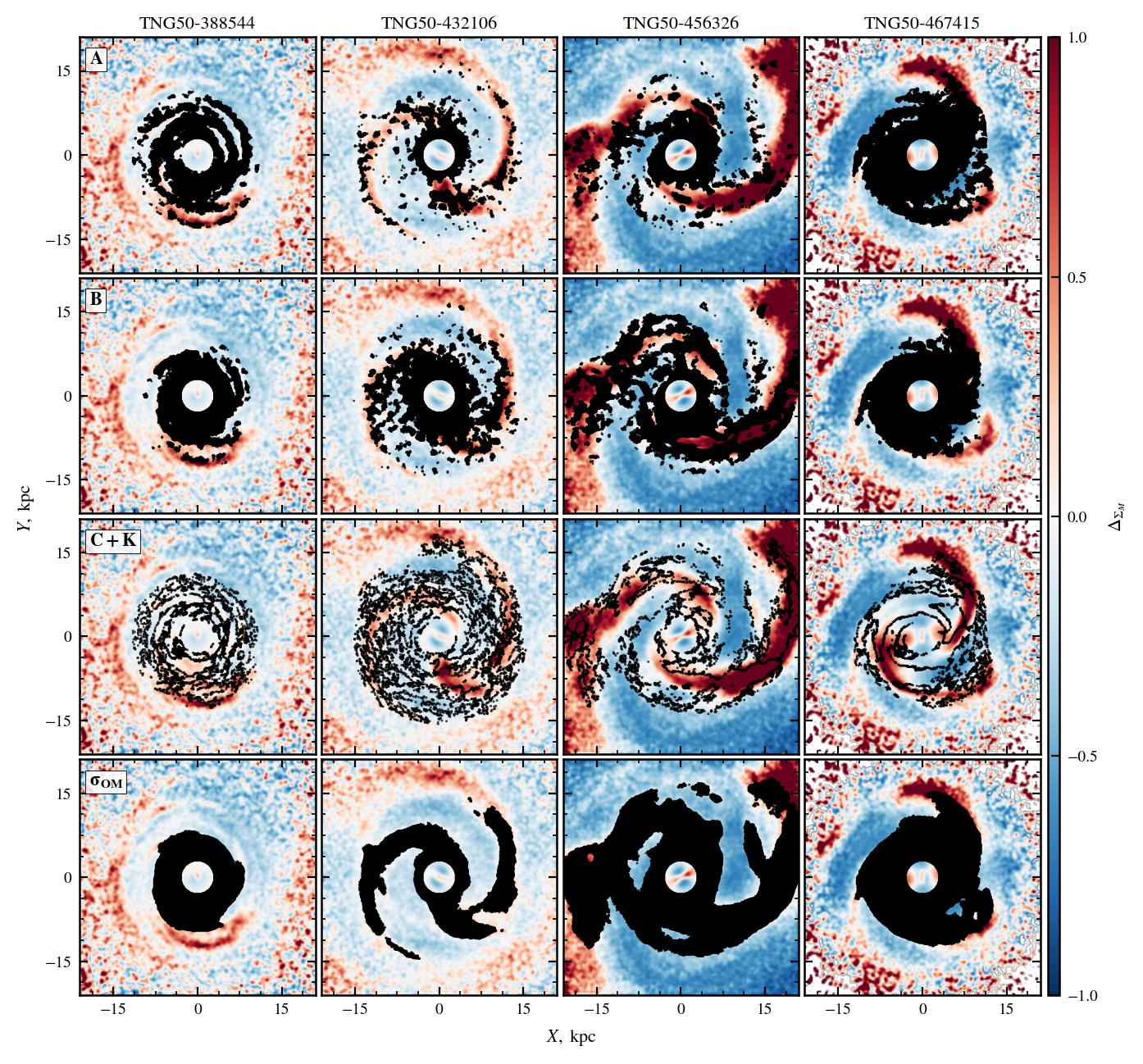}
\caption{
Comparison of the kinematically selected centroids with the stellar-mass overdensity field, $\Delta_{\Sigma_M}$, for the four TNG50 galaxies. Black markers indicate centroids selected according to the adopted kinematic criteria and their uncertainties. The rows correspond to (A) $|M_{12}^{+}|/\sigma(M_{12}^{+})>4$; (B) $|\omega_3|/\sigma(\omega_3)>4$; (C+K) the joint criterion $|M_{11}^{+}|<1~\mathrm{km\,s^{-1}\,kpc^{-1}}$ and $|M_{11}^{+}|/\sigma(M_{11}^{+})<2$; and (bottom) the total Ogorodnikov--Milne uncertainty criterion $\sigma_{\rm OM}<14~\mathrm{km\,s^{-1}\,kpc^{-1}}$. The background colour represents the stellar-mass overdensity field, with positive values indicating an excess of stellar surface density relative to the azimuthally averaged radial profile.
}
\label{fig:overdensity_kinematics}
\end{figure*}

\begin{figure}
\centering
\includegraphics[width=0.9\columnwidth]{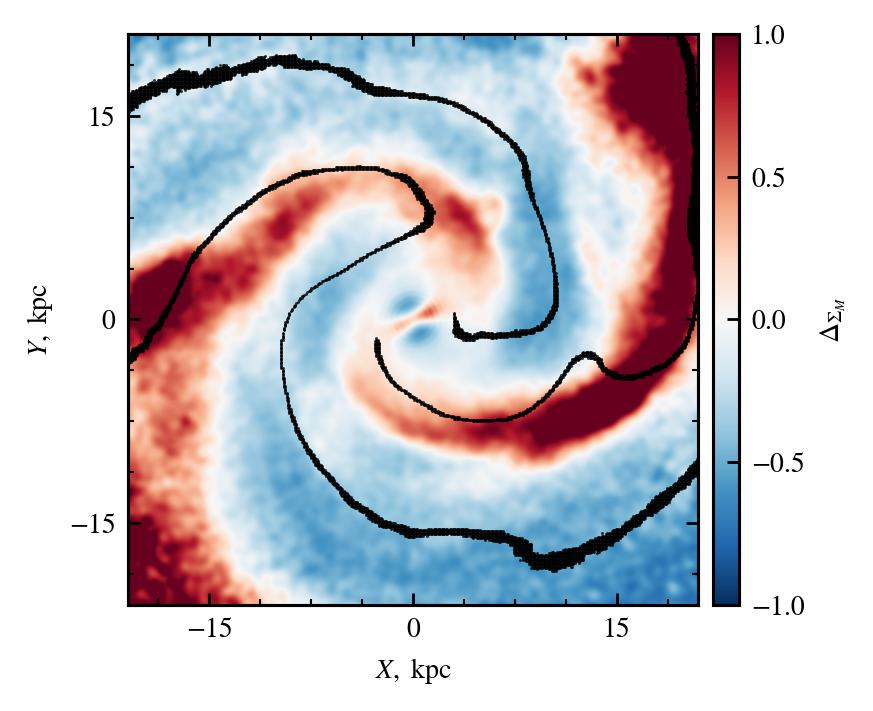}
\caption{Comparison of the kinematically selected centroids with the stellar-mass overdensity field, $\Delta_{\Sigma_M}$, for TNG50-456326. Black markers indicate centroids satisfying the kinematic criterion $|M_{11}^{+}|/\sigma(M_{11}^{+})<2$, determined for a sampling radius of \(r_{\rm loc}=5~\mathrm{kpc}\).}
\label{fig:spiral_tracers}
\end{figure}

The diagonal components of the symmetric deformation tensor measure the velocity gradients along the three coordinate directions. The tensor is defined by Eq.~(\ref{eq:symmetric-tensor-results}), with its diagonal components given by

\begin{equation}
M_{11}^{+}=\frac{\partial V_x}{\partial x},\qquad
M_{22}^{+}=\frac{\partial V_y}{\partial y},\qquad
M_{33}^{+}=\frac{\partial V_z}{\partial z}. \notag
\label{eq:diagonal-components}
\end{equation}

Positive values correspond to local expansion along the corresponding direction, whereas negative values indicate local contraction.

The spatial distributions of $M_{11}^{+}$, $M_{22}^{+}$, and $M_{33}^{+}$ are shown in Fig.~\ref{fig:M11_M22_M33}. The in-plane components $M_{11}^{+}$ and $M_{22}^{+}$ exhibit pronounced spatial variations, with alternating regions of positive and negative deformation across the stellar discs. The amplitude of these structures varies substantially between galaxies. For example, $M_{11}^{+}$ in TNG50-456326 and TNG50-467415 shows relatively coherent large-scale curved and ring-like patterns, while TNG50-432106 displays a smoother and less defined distribution. The corresponding formal uncertainty maps are shown in Fig.~\ref{fig:M11_M22_M33_error}.

Regions in which the diagonal components of the deformation tensor, $M_{11}^{+}$, $M_{22}^{+}$, and $M_{33}^{+}$, change sign mark transitions between local expansion and contraction along the corresponding coordinate axes. In several galaxies, these sign-changing boundaries form extended features, although their morphology and degree of spatial coherence differ among the systems and between tensor components. This motivates an investigation of their relation to structural features in the stellar discs.

\begin{equation}
\begin{gathered}
M_{12}^{+} =
\frac{1}{2}\left(
\frac{\partial V_x}{\partial y}
+
\frac{\partial V_y}{\partial x}
\right),
\qquad
M_{13}^{+} =
\frac{1}{2}\left(
\frac{\partial V_x}{\partial z}
+
\frac{\partial V_z}{\partial x}
\right), \\[4pt]
M_{23}^{+} =
\frac{1}{2}\left(
\frac{\partial V_y}{\partial z}
+
\frac{\partial V_z}{\partial y}
\right).
\end{gathered}
\notag
\end{equation}

The off-diagonal components of the deformation tensor characterise local shear: $M_{12}^{+}$ describes shear within the stellar disc plane, while $M_{13}^{+}$ and $M_{23}^{+}$ describe shear coupling the in-plane and vertical directions. Their spatial distributions for the selected galaxies are shown in Fig.~\ref{fig:M12_M13_M23}.

The $M_{12}^{+}$ component describes in-plane shear and traces the coupling between the $X$ and $Y$ components of the stellar velocity field. In TNG50-388544 and TNG50-467415, $M_{12}^{+}$ forms pronounced ring-like and curved structures around the galactic centre, whereas TNG50-456326 shows a more asymmetric distribution with an extended region of elevated negative values $M_{12}^{+}$. In contrast, the distribution in TNG50-432106 is dominated by smaller-scale fluctuations, suggesting less spatially coherent in-plane shear.

The $M_{13}^{+}$ component describes shear coupling the $X$ and $Z$ directions. Compared with $M_{12}^{+}$, its maps show less pronounced coherent structures within the inner disc. In TNG50-388544 and TNG50-467415, the inner regions are relatively uniform, with values generally close to zero. TNG50-432106 and TNG50-456326 exhibit more heterogeneous distributions, with localised positive and negative features. In particular, TNG50-456326 contains several distinct regions of opposite sign rather than a single coherent large-scale asymmetry.

The $M_{23}^{+}$ component describes shear coupling the $Y$ and $Z$ directions. In TNG50-388544, positive values form an incomplete arc around the inner disc, bordered in places by regions of negative values. TNG50-456326 exhibits several extended positive features, particularly at $Y<0$. TNG50-432106 shows weaker, more dispersed variations, whereas the inner disc of TNG50-467415 remains comparatively uniform, with values close to zero. Both $M_{13}^{+}$ and $M_{23}^{+}$ display stronger small-scale variations towards the outskirts of TNG50-388544 and TNG50-467415. The statistical significance of these variations should be assessed using the corresponding uncertainty maps shown in Fig.~\ref{fig:M12_M13_M23_error} before assigning them a physical interpretation.

The in-plane shear component $M_{12}^{+}$ exhibits the clearest large-scale coherent structures in several galaxies, whereas $M_{13}^{+}$ and $M_{23}^{+}$ are generally more irregular and dominated by smaller-scale variations. A physical interpretation in terms of specific spiral arms or other morphological features requires direct comparison with the stellar-density distribution, presented in Sect.~\ref{sec:spiral-pattern}.

\subsection{Antisymmetric velocity-gradient tensor and local rotation}
\label{sec:rotation-results}

The antisymmetric part of the velocity-gradient tensor describes the local rotational component of the stellar velocity field. Its three independent components can be expressed through the local-rotation vector$\boldsymbol{\omega}=(\omega_1,\omega_2,\omega_3)$ as

\begin{equation}
M^{-} =
\begin{pmatrix}
0 & -\omega_3 & \omega_2 \\
\omega_3 & 0 & -\omega_1 \\
-\omega_2 & \omega_1 & 0
\end{pmatrix}.
\end{equation}

The components $\omega_1$, $\omega_2$, and $\omega_3$ describe local rotation about the $X$, $Y$, and $Z$ axes, respectively. The first two combine in-plane gradients of the vertical velocity with vertical gradients of the in-plane velocities. They therefore describe the rotational component of the local velocity field associated with motions perpendicular to the disc plane. The component $\omega_3$ characterises the local rotation within the disc plane and includes the contribution from the general rotation of the disc. The spatial distributions of these components are shown in Fig.~\ref{fig:omega1_omega2_omega3}. The $\omega_1$ maps display distinct extended features in TNG50-388544 and TNG50-456326. TNG50-432106 exhibits weaker and more dispersed variations.

The $\omega_2$ distributions are relatively smooth in the inner regions of TNG50-388544 and TNG50-467415. In contrast, TNG50-432106 contains an extended negative feature at $Y<0$, while TNG50-456326 shows several localised positive regions, particularly at $Y>0$, interspersed with negative regions. Both $\omega_1$ and $\omega_2$ exhibit strong small-scale variations towards the outskirts of TNG50-388544 and TNG50-467415. The corresponding uncertainty maps are presented in Fig.~\ref{fig:omega1_omega2_omega3_error}.

The $\omega_3$ maps show the clearest common large-scale pattern: all four galaxies contain a prominent central region of negative values, consistent with the dominant sense of disc rotation. In TNG50-388544, this region is surrounded by approximately concentric, ring-like features. TNG50-456326 displays a more asymmetric pattern, with extended negative arcs separated by regions of lower or positive values. In TNG50-432106, the negative central region is more compact, and the outer distribution lacks the pronounced arcs visible in TNG50-388544 and TNG50-456326. Thus, while the central $\omega_3$ signal largely reflects the common rotational structure of the discs, the surrounding patterns differ substantially between galaxies.

\subsection{Kinematic tracers of spiral structure: correspondence with stellar overdensities}
\label{sec:spiral-pattern}

Investigating the spiral structure of the Milky Way using different types of tracers remains an important area of Galactic research \citep{Popova2005, Xu2013, Reid2019, Hao2021, Hou2021}. Following the general approach proposed by \citet{Denyshchenko2024}, we consider the diagonal deformation component $M_{11}^{+}$, the in-plane shear component $M_{12}^{+}$, and the rotation component $\omega_{3}$. These quantities are related to the generalised Oort constants \citep{Chandrasekhar1942,Ogorodnikov1965,Fedorov2023,Akhmetov2024}. In the adopted convention, $M_{12}^{+}=A$, $\omega_{3}=B$, and $M_{11}^{+}=C+K$.

Fig.~\ref{fig:overdensity_kinematics} compares the spatial distributions of centroids selected using three kinematic criteria with the stellar-mass overdensity field $\Delta_{\Sigma_M}$ described in Sect.~\ref{sec:selected_galaxies}. The criteria are labelled A, B, and C according to their relation to the generalised Oort constants. The rows correspond to $|M_{12}^{+}|/\sigma(M_{12}^{+})>4$ (A), $|\omega_3|/\sigma(\omega_3)>4$ (B), and the joint condition $|M_{11}^{+}|<1~\mathrm{km\,s^{-1}\,kpc^{-1}}$ and $|M_{11}^{+}|/\sigma(M_{11}^{+})<2$ (C+K). An additional row, labelled $\sigma_{\rm OM}$, shows centroids with comparatively small overall uncertainties in the O--M parameters. Here, $\sigma_{\rm OM}$ is defined as the quadrature sum of the uncertainties of all nine parameters, and we adopt the criterion $\sigma_{\rm OM}<14~\mathrm{km\,s^{-1}\,kpc^{-1}}$.

The degree of correspondence varies between galaxies. For this comparison, we apply a uniform set of criteria across the sample. We restrict the analysis to $3<R<30~\mathrm{kpc}$ and require a minimum of 500 stellar particles in each sampling region. The same thresholds on the kinematic parameters and their uncertainties are then applied to all four galaxies. However, each galaxy exhibits distinct structural and kinematic features. Improving the correspondence between the kinematically selected features and stellar overdensities may therefore require galaxy-specific choices of these criteria, taking into account galaxy size and morphology.

For example, in TNG50-456326, which exhibits the most clearly defined spiral structure among the galaxies shown, the best correspondence among the tested configurations was obtained with a sampling radius of $r_{\rm loc}=5~\mathrm{kpc}$, a grid spacing of $\Delta_{\rm grid}=0.1~\mathrm{kpc}$, and the criterion $|M_{11}^{+}|/\sigma(M_{11}^{+}) < 2$. The sensitivity of this selection to the adopted local fitting radius is examined in Appendix~\ref{app:rloc_sensitivity}.  At smaller radii, the selected centroids form numerous fragmented and small-scale kinematic features, making it more difficult to distinguish extended spiral-arm segments. As the sampling radius increases, the selection becomes more restrictive, reducing the number of selected centroids while producing narrower and more spatially coherent kinematic sequences. At  $r_{\rm loc}=5~\mathrm{kpc}$, these sequences provide a clearer tracing of the extended spiral structures identified in the stellar-mass overdensity field. This suggests that the characteristic scale of the local fitting region may need to be adjusted to the spatial scales of the structures being traced.Figure~\ref{fig:spiral_tracers} compares the spatial distribution of centroids selected using this $M_{11}^{+}$ with the stellar-mass overdensity field.

For the subsequent spiral fitting, the selected centroids were divided into four spatially coherent groups, labelled I--IV. Each group was fitted independently with a logarithmic spiral, $\ln R=k\phi+b$, using an iterative $3\sigma$-clipping procedure to exclude centroids with large deviations from the fitted relation. The pitch angle is given by $i=\arctan k$. To define the radial normalisation consistently, each fitted spiral was extrapolated inward to $R=1~\mathrm{kpc}$, and the first $\phi=0$ direction encountered when following the spiral outward was adopted as the origin of the first winding. Thus, $R_0=e^b$ is the spiral radius at $\phi=0$. The corresponding best-fitting parameters are summarised in Table~\ref{tab_spiral}.  The procedures for centroid selection, determination of the spiral parameters, and estimation of their associated uncertainties follow the methodology described by \citet{Denyshchenko2024}.

\begin{figure}
\centering
\includegraphics[width=0.9\columnwidth]{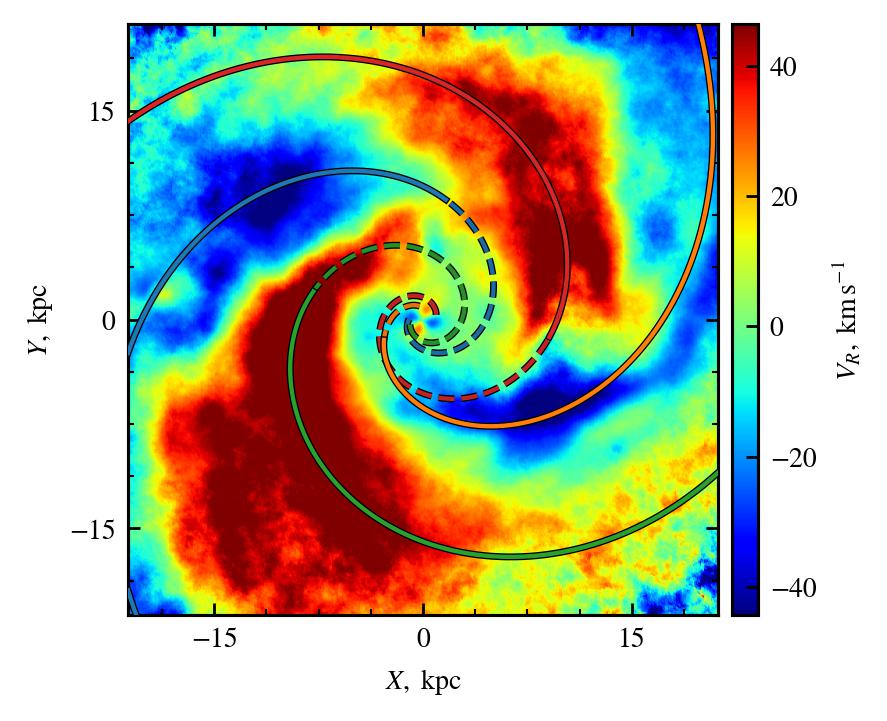}
\caption{Logarithmic spiral fits for the TNG50-456326 galaxy overlaid on the radial-velocity field, $V_R$. The four fitted spiral segments are shown as I (green), II (blue), III (red), and IV (orange). The corresponding fitted spiral segments are shown by solid curves, while the dashed parts indicate their inward extrapolation.  The corresponding spiral-fit parameters are listed in Table~\ref{tab_spiral}.}
\label{fig:spiral_fits_vr}
\end{figure}

\begin{figure}
\centering
\includegraphics[width=0.9\columnwidth]{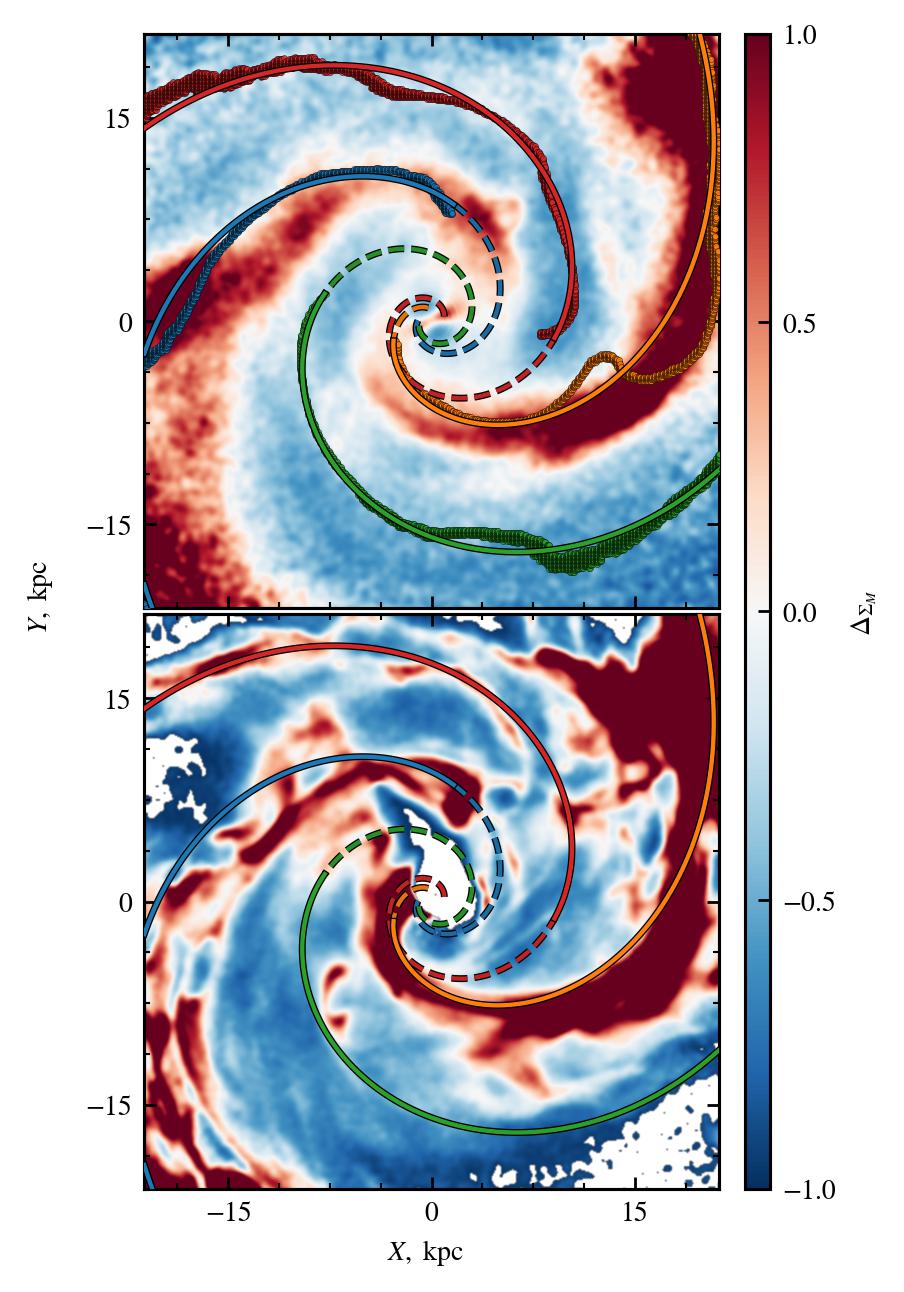}
\caption{Logarithmic spiral fits for the TNG50-456326 galaxy overlaid on the stellar- and gas-mass overdensity fields, $\Delta_{\Sigma_M}$ (top and bottom, respectively). In the upper panel, coloured points indicate the selected stellar centroids assigned to the four spiral segments: I (green), II (blue), III (red), and IV (orange). The corresponding fitted spiral segments are shown by solid curves, while the dashed parts indicate their inward extrapolation. The corresponding spiral-fit parameters are listed in Table~\ref{tab_spiral}.}
\label{fig:spiral_fits_stellar_gas}
\end{figure}

For TNG50-456326, we additionally constructed a gas-mass overdensity map using the same spatial selection, grid, smoothing parameters, and radial-background procedure as for the stellar-mass overdensity field described in Sect.~\ref{sec:selected_galaxies}. The gaseous component provides an additional tracer of spiral morphology and allows the fitted structures to be compared with both the stellar and gas distributions \citep{DobbsBaba2014,Hou2021}. Figs.~\ref{fig:spiral_fits_vr} and \ref{fig:spiral_fits_stellar_gas} compare the four fitted logarithmic spirals in TNG50-456326 with the radial-velocity field and the stellar- and gas-mass overdensity fields, respectively. Their morphology suggests two principal arms and two secondary branches, qualitatively resembling proposed Milky Way models \citep{Reid2009, Robitaille2012}. The fitted curves wind towards the inner disc, suggesting a possible connection with the ends of a central bar. This interpretation requires independent confirmation, particularly because the innermost disc was excluded from the tracer selection. The fitted spiral segments also show an apparent correspondence with local extrema in the radial-velocity field.

\begin{table}
\caption{Parameters of the logarithmic spiral fits for the four spatially coherent spiral segments identified in TNG50-456326. Here, $N$ is the number of centroids used in each fit, $R_0$ is the spiral radius at $\phi=0$ on the adopted first winding, and $i$ is the corresponding pitch angle.}
\label{tab_spiral}
\centering
\small
\setlength{\tabcolsep}{4pt}
\begin{tabular}{lccc}
\hline\hline
Segment & $N$ & $R_0$ (kpc) & $i$ (deg) \\
\hline
I & 3432 & $2.796 \pm 0.015$  & $20.275 \pm 0.051$ \\
II   & 2135 & $4.543 \pm 0.017$   & $25.413 \pm 0.054$ \\
III  & 4063 & $9.651 \pm 0.022$   & $20.880 \pm 0.050$ \\
IV  & 2317 & $17.205 \pm 0.03$  & $32.48 \pm 0.089$ \\
\hline
\end{tabular}
\end{table}

\section{Conclusions}
\label{sec:conclusions}

In this work, we investigated whether local O--M parameters can be used to identify spiral and other non-axisymmetric structures in four Milky Way+Andromeda galaxies from the TNG50-1 simulation at $z=0$. We focused on exploring additional kinematic criteria and assessing their correspondence with stellar-mass overdensities defined independently of the kinematic selection. 

Local O–M parameters provide complementary information to centroid velocities. The in-plane shear component ($M_{12}^{+}$) reveals extended coherent structures, whereas components involving vertical motions generally show less spatially coherent patterns. These results demonstrate the potential of local velocity gradients to characterise kinematic structures that are not readily identified from centroid velocities alone.

 In TNG50-456326, the selected centroids trace extended segments of the prominent overdense spiral arms, whereas in TNG50-432106 and TNG50-467415 the agreement is more partial and varies along the arms. For TNG50-388544, the selected structures are concentrated in the inner disc, where their association with individual overdense features is less evident. The selections based on $A$ and $B$ generally cover broader, more contiguous regions, while those based on $C+K$ delineate narrower, more fragmented curved sequences. The broad spatial overlap with the $\sigma_{\rm OM}$ selection further highlights the need to account for parameter uncertainties when interpreting these patterns.

The reduced number of selected centroids at large Galactocentric radii may reflect limitations in the local fits. The lower number of stellar particles available in the outer regions can lead to less constrained parameter estimates, larger uncertainties, and greater scatter in the measured velocity field. Consequently, the lack of selected centroids at large radii does not necessarily indicate an absence of spiral structure. 

The spatial correspondence between selected centroids and density structures may depend on the adopted fitting scale. In TNG50-456326 and TNG50-467415, some selected centroids lie near the boundaries of stellar overdensities, where the density gradient is relatively steep. One possible explanation is a mismatch between the radius of the local spherical regions used for the O–M fits and the characteristic width of the spiral structures, which may be substantially larger.

For TNG50-456326, a galaxy-specific selection enabled us to fit four logarithmic spiral branches. Their morphology and fitted parameters, together with the behaviour of the local kinematic fields, suggest similarities to the Milky Way, including the kinematic patterns mapped by \citet{Fedorov2023}. These similarities identify TNG50-456326 as a promising candidate for a kinematic analogue of our Galaxy.

The apparent spatial correspondence between overdense spiral segments and local extrema of the radial velocity field may reflect coherent radial streaming motions associated with non-axisymmetric perturbations. A complementary connection between spiral structure and stellar orbital properties was reported by \citet{Palicio2023}, who identified low-radial-action structures coinciding with several Galactic spiral-arm segments. Furthermore, \citet{Hamilton2026} showed that transient spiral perturbations can generate a wave-like dependence of mean radial velocity on angular momentum resembling that observed in the Galactic disc. These studies provide a dynamical context for interpreting the similarities identified here, without establishing that the same mechanisms operate in both galaxies.

The value of TNG50-456326 as a candidate analogue lies in the possibility of tracing its assembly history and dynamical evolution within the physical framework of the simulation. This provides an opportunity to investigate how the formation and evolution of spiral structure, internal perturbations, and interactions with satellites are reflected in the local kinematic parameters. We also compared the stellar structures identified in this analysis with the gas distribution, providing an additional perspective on the spatial relationship between the stellar and gaseous components. Establishing these connections could help constrain the evolutionary processes underlying similar patterns observed in the Milky Way.

Future work will therefore focus on developing and testing additional kinematic criteria and applying them to multiple simulation snapshots. Tracking the selected features over time will allow us to examine their relationship with spiral-arm evolution, stellar motions, and external perturbations, and to assess the persistence of the similarities between TNG50-456326 and the Milky Way. A further step will be to extend the criteria tested here to Gaia observations of the Milky Way. This will require appropriate stellar samples and an assessment of the effects of the observer's position within the disc, spatial selection functions, extinction, and measurement uncertainties. Such an analysis will test the applicability of these diagnostics to observational data and their usefulness for investigating the kinematics and dynamics of Galactic spiral structure.


\section*{Acknowledgements}

This work has made use of data from the TNG50 simulation of the IllustrisTNG project, including the publicly available TNG50 Milky Way+Andromeda (MW/M31) galaxy sample. The authors acknowledge the IllustrisTNG collaboration for developing the simulations and for providing public access to the simulation data and the MW/M31-like galaxy sample.

This work is supported by the National Research Foundation of Ukraine, Project No. 2023.03/0188.



\section*{Data availability}

The TNG50 simulation data used in this work, including the Milky Way+Andromeda (MW/M31) galaxy sample, are publicly available through the IllustrisTNG data portal (\url{https://www.tng-project.org/}). The software code
used in this paper can be made available upon request by emailing
the corresponding authors.


\bibliographystyle{mnras}
\bibliography{references}

\appendix

\section{Number of stellar particles in the local kinematic samples}
\label{app:starsAmount}

\begin{figure*}
\centering
\includegraphics[width=0.9\textwidth]{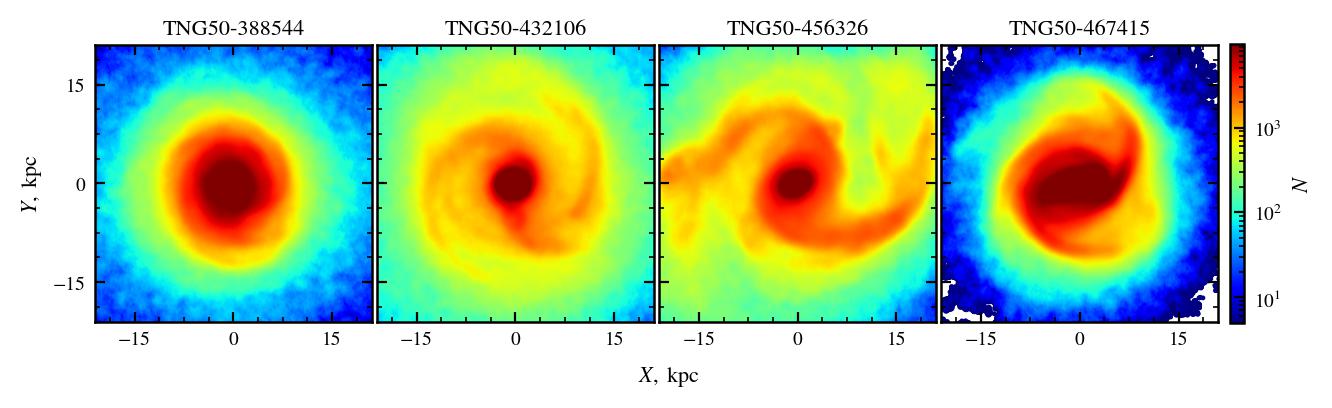}
\caption{
Spatial distribution of the number of stellar particles, $N$, contributing to the local kinematic solutions for the four selected TNG50-1 galaxies. The colour scale is logarithmic. Each panel corresponds to one galaxy listed in Table~\ref{tab:galaxies_r30}.
}
\label{fig:starsAmount}
\end{figure*}

Figure~\ref{fig:starsAmount} shows the spatial distribution of the number of stellar particles contributing to the local kinematic solutions for the four selected TNG50-1 galaxies. The number of particles generally decreases towards the outer disc and also shows non-axisymmetric variations associated with the stellar structure of each galaxy. These differences in sampling density provide useful context for interpreting the spatial distributions of the uncertainties of the O--M parameters presented in Appendix~\ref{app:parameter_errors}.

\section{Uncertainties of the local kinematic parameters}
\label{app:parameter_errors}

\begin{figure*}
\centering
\includegraphics[width=0.80\textwidth]{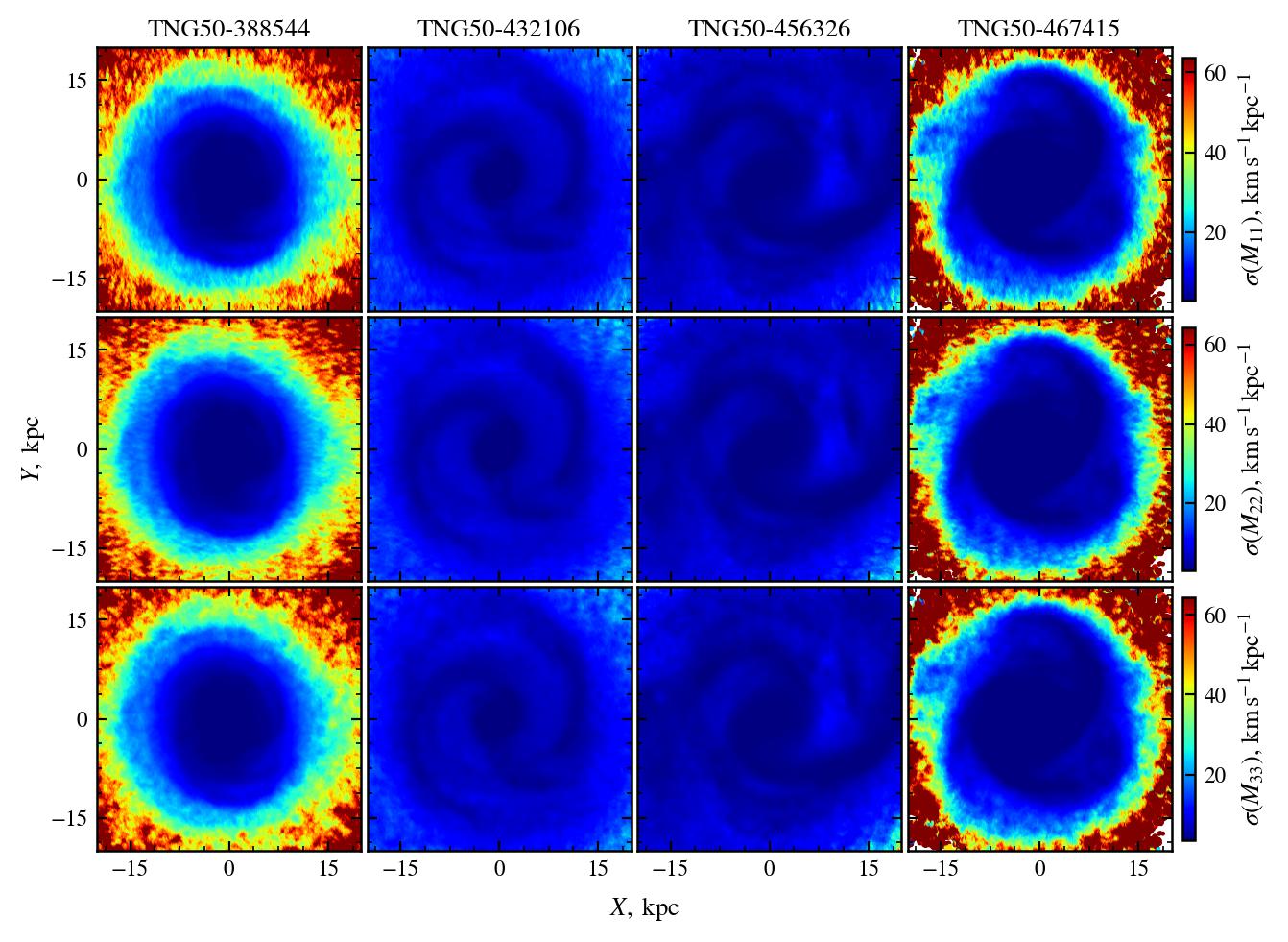}
\caption{
Maps of the uncertainties of the diagonal deformation components $M_{11}^{+}$, $M_{22}^{+}$, and $M_{33}^{+}$ for the selected TNG50-1 galaxies. From top to bottom, the rows show $\sigma(M_{11}^{+})$, $\sigma(M_{22}^{+})$, and $\sigma(M_{33}^{+})$, respectively. Each column corresponds to one galaxy listed in Table~\ref{tab:galaxies_r30}.
}
\label{fig:M11_M22_M33_error}
\end{figure*}

\begin{figure*}
\centering
\includegraphics[width=0.80\textwidth]{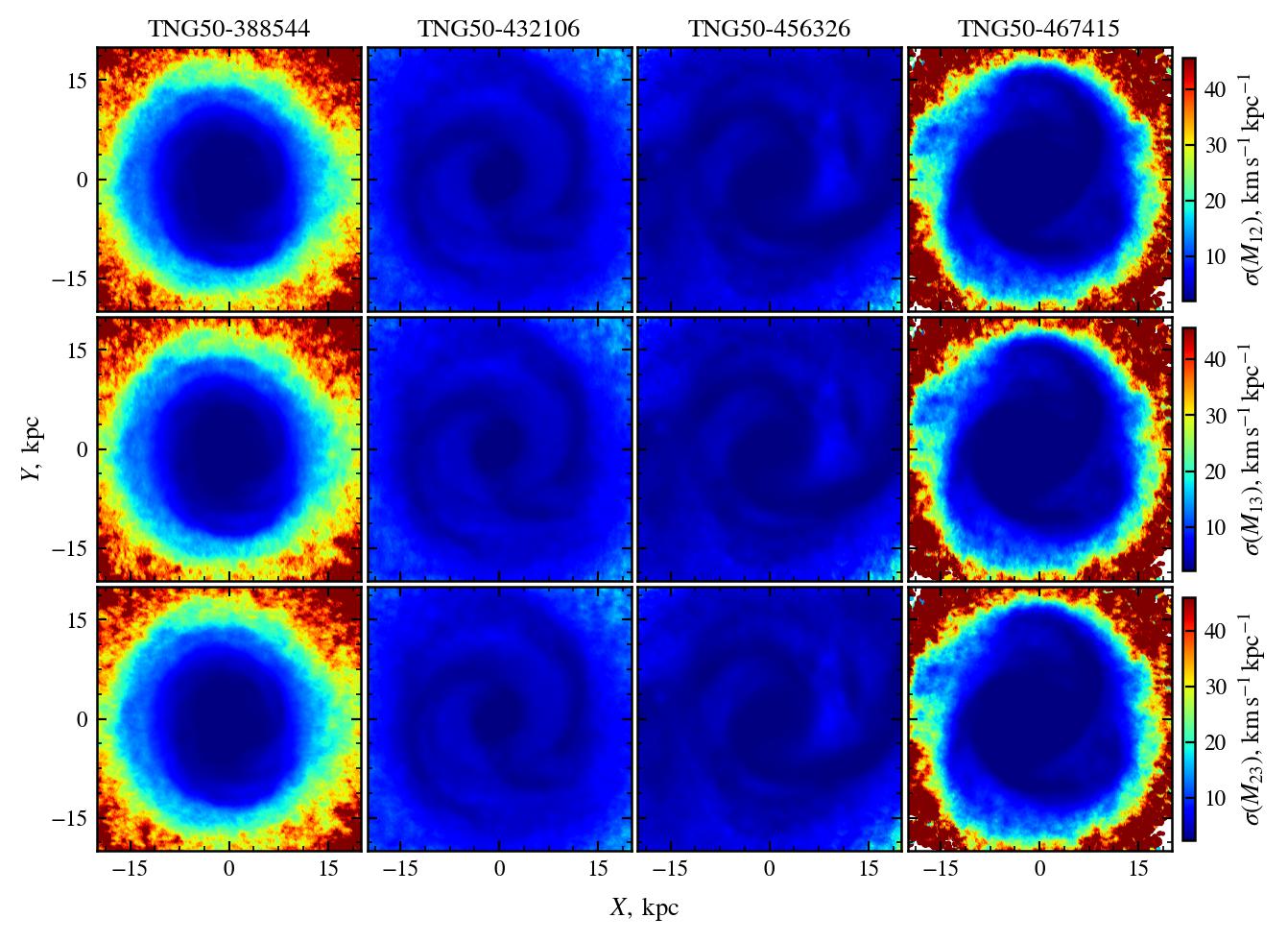}
\caption{
Maps of the uncertainties of the off-diagonal deformation components $M_{12}^{+}$, $M_{13}^{+}$, and $M_{23}^{+}$ for the selected TNG50-1 galaxies. From top to bottom, the rows show $\sigma(M_{12}^{+})$, $\sigma(M_{13}^{+})$, and $\sigma(M_{23}^{+})$, respectively. Each column corresponds to one galaxy listed in Table~\ref{tab:galaxies_r30}.
}
\label{fig:M12_M13_M23_error}
\end{figure*}

\begin{figure*}
\centering
\includegraphics[width=0.80\textwidth]{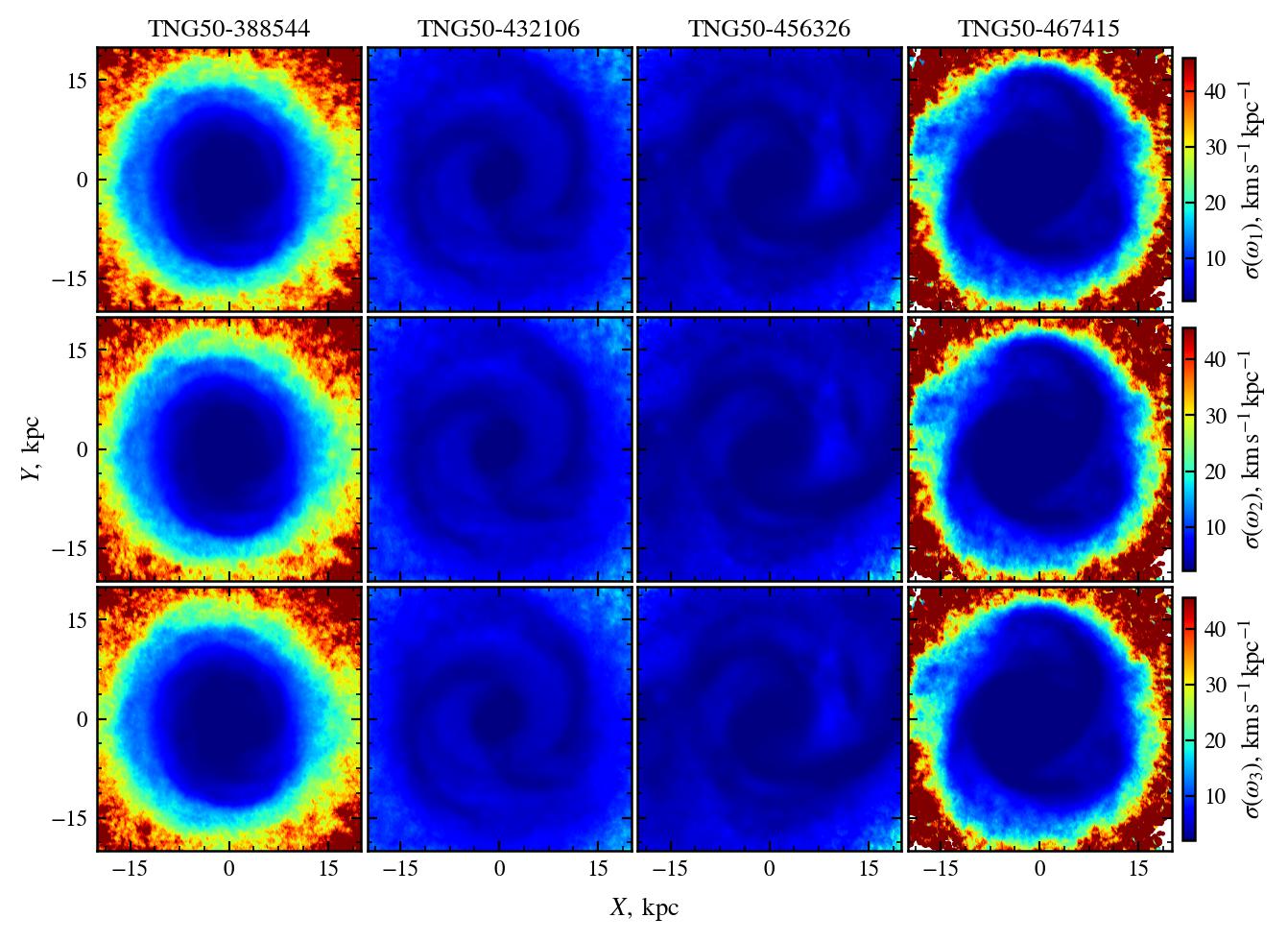}
\caption{
Maps of the uncertainties of the local-rotation components $\omega_1$, $\omega_2$, and $\omega_3$ for the selected TNG50-1 galaxies. From top to bottom, the rows show $\sigma(\omega_1)$, $\sigma(\omega_2)$, and $\sigma(\omega_3)$, respectively. Each column corresponds to one galaxy listed in Table~\ref{tab:galaxies_r30}.
}
\label{fig:omega1_omega2_omega3_error}
\end{figure*}

Figures~\ref{fig:M11_M22_M33_error}--\ref{fig:omega1_omega2_omega3_error} show the spatial distributions of the uncertainties of the local Ogorodnikov--Milne parameters for the selected TNG50-1 galaxies. These maps complement the corresponding parameter distributions presented in the main text and illustrate the spatial variation in the precision of the local kinematic solutions. In general, the uncertainties increase towards the outer parts of the discs, where the stellar-particle sampling becomes sparser, consistent with the distributions shown in Appendix~\ref{app:starsAmount}.

\section{Sensitivity to the local fitting radius}
\label{app:rloc_sensitivity}

The spatial distribution of the kinematically selected centroids depends on the characteristic scale over which the local velocity field is fitted. To examine this dependence, we repeated the $M_{11}^{+}$-based selection for TNG50-456326 using local sampling radii of $r_{\rm loc}=1$, 2, 3, and 5~kpc, while keeping the remaining selection parameters fixed. In all cases, the centroids were selected using the criterion $|M_{11}^{+}|/\sigma(M_{11}^{+})<2$ and compared with the stellar-mass overdensity field, $\Delta_{\Sigma_M}$.

\begin{figure*}
\centering
\includegraphics[width=0.9\textwidth]{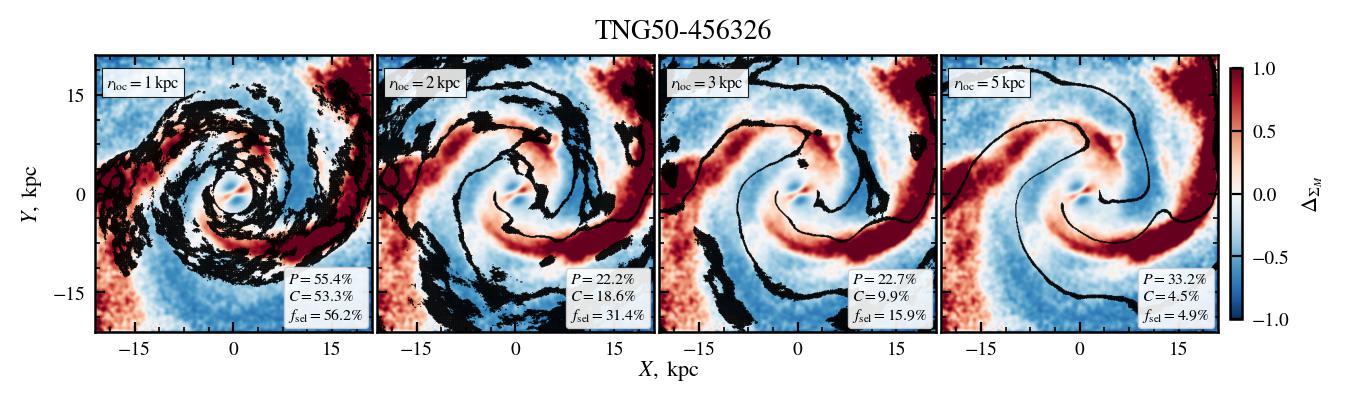}
\caption{Comparison of the kinematically selected centroids with the stellar-mass overdensity field, $\Delta_{\Sigma_M}$, for TNG50-456326 at different local sampling radii, $r_{\rm loc}$. Black markers indicate centroids satisfying the kinematic criterion $|M_{11}^{+}|/\sigma(M_{11}^{+})<2$. The panels correspond to $r_{\rm loc}=1$, 2, 3, and 5~kpc. The quantities $P$, $C$, and $f_{\rm sel}$ denote the purity, completeness, and fraction of selected centroids, respectively.}
\label{fig:rloc_sensitivity}
\end{figure*}

Figure~\ref{fig:rloc_sensitivity} illustrates how the spatial distribution of the selected centroids changes with $r_{\rm loc}$. We quantify the correspondence using the purity $P=N_{\rm int}/N_{\rm sel}$, the completeness $C=N_{\rm int}/N_{\rm ref}$, and the selected fraction $f_{\rm sel}=N_{\rm sel}/N_{\rm sample}$, reported in per cent. Here, $N_{\rm sel}$ is the number of kinematically selected centroids, $N_{\rm ref}$ is the number located in regions with $\Delta_{\Sigma_M}>0$, $N_{\rm int}$ is their intersection, and $N_{\rm sample}$ is the total number of analysed centroids.

Increasing $r_{\rm loc}$ reduces both the completeness and the fraction of selected centroids, from $C=53.3\%$ and $f_{\rm sel}=56.2\%$ at $r_{\rm loc}=1$~kpc to $C=4.5\%$ and $f_{\rm sel}=4.9\%$ at $r_{\rm loc}=5$~kpc, consistent with stronger spatial averaging of the local velocity-gradient field at larger fitting scales. At small $r_{\rm loc}$, the selected centroids cover a larger fraction of the disc and include more fragmented small-scale features. In contrast, the $r_{\rm loc}=5$~kpc solution is more restrictive and isolates narrower, spatially coherent sequences associated with the prominent spiral structure of TNG50-456326. This motivated the adoption of $r_{\rm loc}=5$~kpc for the subsequent identification and fitting of individual spiral segments.

\bsp
\label{lastpage}
\end{document}